\documentclass[a4paper, amsfonts, amssymb, amsmath, reprint, showkeys, nofootinbib, oneside]{revtex4-1}
\usepackage[left=2cm, right=2cm, top=2cm, bottom=2cm]{geometry}
\usepackage[english]{babel}
\usepackage[utf8]{inputenc}
\usepackage[colorinlistoftodos, color=green!40, prependcaption]{todonotes}
\usepackage{amsthm}
\usepackage{mathtools}
\usepackage{physics}
\usepackage{xcolor}
\usepackage{graphicx}
\usepackage{adjustbox}
\usepackage{placeins}
\usepackage[T1]{fontenc}
\usepackage{lipsum}
\usepackage{csquotes}
\usepackage[pdftex, pdftitle={Article}, pdfauthor={Author}]{hyperref} % For hyperlinks in the PDF

\usepackage[utf8]{inputenc}  % 支持UTF-8编码
\usepackage[T1]{fontenc}     % 优化字符输出

\begin{document}
\title{Exploiting scattering-based point spread functions for snapshot 5D and modality-switchable lensless imaging}

\author{Ze Zheng\(^{1,2}\)}
\author{Baolei Liu\(^{3}\)}
\email[Correspondence e-mail: ]{liubaolei@buaa.edu.cn}
\author{Jiaqi Song\(^{1,4}\)}
\author{Muchen Zhu\(^{1}\)}
\author{Yao Wang\(^{1}\)}
\author{Menghan Tian\(^{1}\)}
\author{Ying Xiong\(^{1}\)}
\author{Zhaohua Yang\(^{3}\)}
\author{Xiaolan Zhong\(^{1}\)}
\author{David McGloin\(^{5}\)}
\author{Fan Wang\(^{1}\)}

\affiliation{\(^{1}\)School of Physics, Beihang University, Beijing 100191, China\\ 
\(^{2}\)State Key Laboratory of Photonics and Communications, Institute for Quantum Sensing and Information Processing, Shanghai Jiao Tong University, Shanghai 200240, P.R. China\\ 
\(^{3}\)School of Instrumentation and Optoelectronics Engineering, Beihang University, Beijing 100191, China\\
\(^{4}\)Institute of Physics, Chinese Academy of Sciences, Beijing 100190, China\\ 
\(^{5}\)School of Natural and Computing Science, University of Aberdeen, King’s College, Aberdeen, AB24 3FX, UK}

\date{\today} % Leave empty to omit a date

\begin{abstract}
Snapshot multi-dimensional imaging offers a promising alternative to traditional low-dimensional imaging techniques by enabling the simultaneous capture of spatial, spectral, polarization, and other information in a single shot for improved imaging speed and acquisition efficiency. However, existing snapshot multi-dimensional imaging systems are often hindered by their large size, complexity, and high cost, which constrain their practical applicability. In this work, we propose a compact lensless diffuser camera for snapshot multi-dimensional imaging (Diffuser-mCam), which can reconstruct five-dimensional (5-D) images from a single-shot 2D recording of speckle-like measurement under incoherent illumination. By employing both the scattering medium and the space-division multiplexing strategy to extract high-dimensional optical features, we show that the multi-dimensional data (2D intensity distribution, spectral, polarization, time) of the desired light field can be encoded into a snapshot speckle-like pattern via a diffuser, and subsequently decoded using a compressed sensing algorithm at the sampling rate of 2.5\%, eliminating the need for multi-scanning processes. We further demonstrate that our method can be flexibly switched between 5D and selectively reduced-dimensional imaging, providing an efficient way of reducing computational resource demands. Our work presents a compact, cost-effective, and versatile framework for snapshot multi-dimensional imaging and opens up new opportunities for the design of novel imaging systems for applications in areas such as medical imaging, remote sensing, and autonomous systems.
\end{abstract}

\keywords{snapshot imaging, multi-dimensional imaging, scattering, lensless imaging, compact imaging system}

\maketitle

\section{Introduction} \label{sec1:introduction}
    % \lipsum[2-3]
    \begin{figure*}[htbp]
    \centering
    \includegraphics[width=\textwidth]{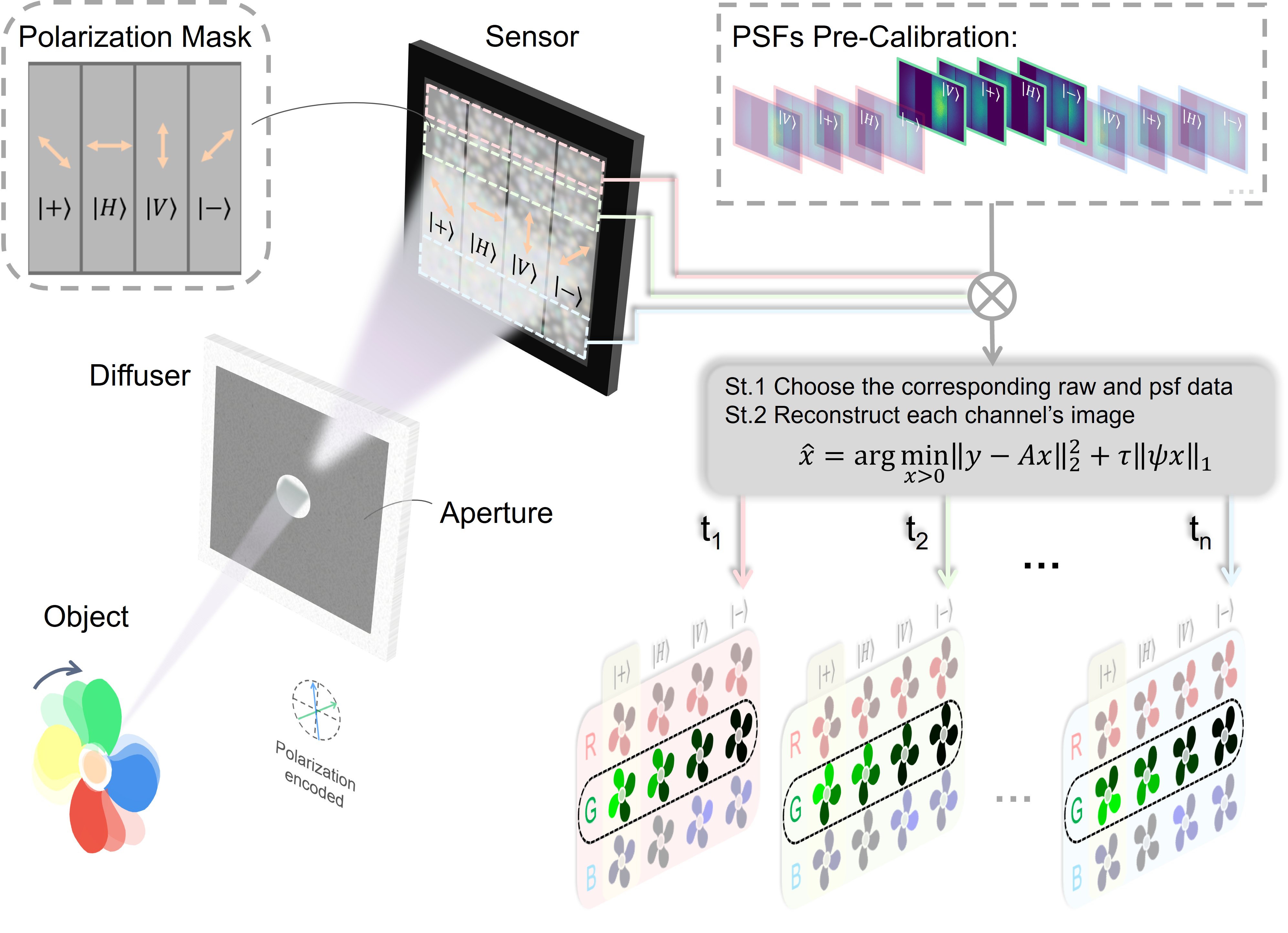}
    \caption{Concept and principle of the proposed compact snapshot multi-dimensional imaging method with a lensless diffuser camera (Diffuser-mCam). The front-end perception part contains a diffuser and a customized polarization mask to enhance the encoding of the input dynamic high-dimensional light field. An aperture is placed behind the diffuser to block large angles of incident light to ensure that the entire PSF and Raw data are captured by the sensor. The polarization mask is custom-made of 4 thin-film polarizers with different linear polarization directions: ${\vert H\rangle}$ , ${\vert+\rangle}$ , ${\vert V\rangle}$ and ${\vert-\rangle}$. The CMOS sensor snapshots an encoded raw image of the input light field in the rolling-shutter mode. For pre-calibration, a 75 $\mu$m pinhole was placed in the object plane instead of the objects to generate 12 spectral-polarization PSFs with four different polarization directions and three different wavelengths (red, green, and blue). The 12 spectral-polarization joint encoded PSFs of Diffuser-mCam are shown in the gray dashed box in the top-right corner. By under-sampling different small regions (as shown in the sensor plane) parallel to the row of raw data and the corresponding PSFs, the multi-dimensional imaging results (2D intensity distribution, spectral, polarization, time) can be reconstructed by the CS algorithm, as shown at the bottom. The green polarization results in the black dashed ovals are related to the highlighted four PSFs.}
    \label{fig:1}  
    \end{figure*}
    
    Multi-dimensional optical imaging systems have seen significant advancements over the past decade, which can acquire high-dimensional information such as 2D spatial distribution, wavelength, time, and polarization for more comprehensive sensing of the physical world. Traditional multi-dimensional imaging systems typically rely on scanning mechanisms, which can be time-consuming and cumbersome. In contrast, recently developed snapshot multi-dimensional imaging systems, such as snapshot Image Mapping Spectrometer (IMS)\cite{gao2010snapshot}, code aperture snapshot spectral imaging (CSSCI)\cite{arguello2012rank}, and sequentially timed all-optical mapping photography (STAMP)\cite{nakagawa2014sequentially}, enable parallel acquisition mechanisms to capture high-dimensional data cubes. Specifically, IMS achieves parallel acquisition by optically slicing the image into multiple sub-images and redirecting them onto different regions of a detector, each corresponding to a different spectral band. CASSI uses a coded aperture and dispersive optics to spatially and spectrally modulate the scene simultaneously across the entire detector array. STAMP, on the other hand, relies on ultrafast optical shutters and time-staggered beam replicas to encode temporal information into a single-shot 2D measurement. These mechanisms allow spatial, spectral, and temporal dimensions to be encoded in parallel, significantly improving acquisition speed and making them particularly suitable for dynamic scenes. The fusion of the optical system and the advanced computational algorithm enhances the reconstruction of high-dimensional data and reduces the overall measurement time\cite{wang2025computational}. Snapshot multi-dimensional imaging techniques have been widely applied in various fields, such as plenoptic imaging\cite{hua2022ultra, lin2022end, park2020snapshot, joshi2025interferenceless}, hyperspectral imaging\cite{yako2023video, bian2024broadband}, and polarimetric imaging\cite{rubin2019matrix}. However, the existing systems still require the utilization of precision instruments, including lenses, beam splitters, meta-surfaces, and costly detectors\cite{gao2016review}. They also face challenges in terms of size, weight, complexity, and cost, which restrict their implementation in practical applications. 
    
  Scattering media have demonstrated their practical potential in areas such as photonic computing\cite{leonetti2021optical,pierangeli2021scalable,wang2024large}, optical reservoir computing\cite{ding2024optoelectronic}, and optical encryption\cite{bian2024large}. By introducing random scattering into optical systems, complex light modulation that encodes optical information in a highly efficient manner can be created. The ability to manipulate light through scattering has made it an ideal candidate for its utilization in compact snapshot multi-dimensional imaging systems, where it can facilitate the encoding of multi-dimensional information without the need for large, bulky optical components. The lensless 3D diffuser camera\cite{antipa2017diffusercam} has been demonstrated, which extracts depth information using point-spread functions (PSFs) engineering of a diffuser. Recent research into scattering media-based high-dimensional imaging has demonstrated numerous applications, including: light field imaging\cite{cai2020lensless}, polarization imaging\cite{elmalem2021lensless,baek2022lensless,pierangeli2024deep}, multi/hyper-spectral\cite{sahoo2017single,li2019single,li2022hyperspectral,kim2023aperture,malone2023diffuserspec,monakhova2020spectral,liu2021single}, in-vivo high-contrast imaging\cite{adams2022vivo,wu2024mesoscopic}, temporal compressive imaging\cite{zheng2024temporal,antipa2019video,weinberg2020100}, sensing\cite{wu2024multiplexed,zhu2023harnessing}, super-resolution imaging\cite{sun2024overcoming,tian2024cfza}. In addition, some special scattering mediums, such as meta-surface mask\cite{shen2023monocular}, liquid-crystal diffuser\cite{lei2023snapshot}, and reconfigurable particle assembly masks\cite{miller2020particle,zhang2025lensless}, offer potential potential in programmable imaging applications.
    
    In this work, we propose a compact snapshot multi-dimensional imaging method assisted by a diffuser (Diffuser-mCam) to extract 5-dimensional information (2D intensity distribution, spectral, polarization, time) of the light field from a single measurement. The proposed system, consisting of a diffuser, a polarization mask, and a commercial bare CMOS camera, is designed as an on-chip imaging system, which is both compact (27 mm × 27 mm × 7 mm) and light-weight (~6 g). By employing the diffuser and the space-division multiplexing strategy, Diffuser-mCam is able to encode the 432 channels of the incoherent multi-dimensional light field into a snapshot speckle-like raw data (i.e., 6048 channels per second) at a sampling rate of 2.5 \%. In the pre-calibration process, a pinhole was put on the object plane, and 4 sets of irrelevant spectral-polarization joint-encoded PSFs (see Experimental Section) of the Diffuser-mCam were separately generated and captured, corresponding to various imaging modalities. By inputting the raw data and the corresponding channel’s PSF, the imaging results of each dimensional channel can be subsequently inverse-solved by a compressed sensing (CS) algorithm. Furthermore, Diffuser-mCam provides an on-demand framework for modality-switchable reconstruction to efficiently conserve computational resources. Based on the forward imaging model and the fact that the calibrated PSFs are weakly correlated with each other, the reconstructions for different channels can be performed separately. Thus, Diffuser-mCam only needs to reconstruct the desired modalities, which increases the sampling rate and improves the reconstruction quality (see the Supporting Information S3). For example, only the polarization images are required to be reconstructed in a classification task for static transparently polarized objects, without reconstructing images for the other dimensions. Diffuser-mCam provides a lensless, compact, and low-cost method to efficiently encode and reconstruct incoherent multi-dimensional optical information, representing a novel multi-dimensional optical imaging scheme.

  %I believe leaving the sections in separate files is more organized, change it if you desire 
\section{Results} \label{sec2:result}
    %\lipsum[4-5]
    \subsection{The principle of Diffuser-mCam}
    Figure 1 illustrates the concept and principle of the proposed Diffuser-mCam. The target is a rotating or moving dynamic object, which has both spectral and polarization information. The linearly polarized basis vectors are defined as ${\vert H\rangle}$ and ${\vert V\rangle}$, where ${\vert H\rangle}$ is the horizontal direction and ${\vert V\rangle}$ is the vertical direction. The diagonally polarized basis vectors are defined as ${\vert+\rangle}$ and ${\vert-\rangle}$, corresponding to the direction of + 45° and - 45°, respectively. The incoherent light field from the target is first scattered by a diffuser. Since the diffuser scatters differently for light with different wavelengths and polarizations, the speckle-like light field on the detection plane is encoded with spectral and polarization characteristics. To further enhance the polarization-based encoding, here we place a polarization mask near the detection sensor to encode the polarization information into different column regions on the sensor plane. As shown in Figure 1, the polarization mask is a customized thin-film polarizer array comprising 4 vertical regions, each with a distinct linear polarization direction: ${\vert H\rangle}$, ${\vert+\rangle}$, ${\vert V\rangle}$, ${\vert-\rangle}$. These regions are arranged along the columns of the sensor. The speckle-like effective raw image (360 × 500 pixels) with 2D grayscale intensity is captured by the CMOS sensor in rolling shutter mode. In rolling-shutter mode, the rows of the sensor begin to expose at different times to record the dynamically changing information of the scattered light field. In this paper, the term “super-row” is used to describe each ten adjacent rows of raw data that record information about the light field within a similar period of time, as shown in the different colored dashed boxes on the sensor plane of Figure 1. Thus, the effective raw data is divided into 36 super-rows (see the Experimental Section). The short row delay time of the sensor decides the temporal resolution of Diffuser-mCam. A total of four sets, comprising 20 weakly correlated PSFs of Diffuser-mCam, were pre-calibrated by a 75 $\mu$m pinhole. The selected 12 spectral-polarization joint encoded PSFs, which correspond to the five-dimensional (2D spatial-temporal-spectral-polarization) imaging modality, are shown in the gray dashed box in the top-right corner of Figure 1. They are denoted by their PSFs index from 1 to 12 in order (PSF Set \#1). The other sets of PSFs contain three spectral PSFs (PSF Set \#3: corresponding to the multi-spectral imaging modality), four white-polarization PSFs (PSF Set \#2: corresponding to the polarization-stock imaging modality), and one white-natural PSF (PSF Set \#4corresponding to the temporal imaging modality) are denoted by the PSFs index from 13 to 15, from 16 to 19, and 20, respectively. The detailed experimental setup is shown in the experimental section.
    
    According to the forward imaging model, the grayscale raw data \emph{I} can be represented by the convolution with the multi-dimensional object \emph{O} and the PSF, within the range of optical memory effect:
    \begin{equation}
        I(x,y,t) = \sum_j O_j(x,y,\lambda,p,t) \cdot \text{PSF}_j(x,y,\lambda,p),
    \end{equation}
    where \emph{I} is the grayscale raw data captured by the CMOS sensor in rolling shutter mode, \emph{O} is the multi-dimensional object, and j is the spectral-polarization PSFs’ index. The multi-dimension of the light field is represented respectively: the 2D intensity distribution (x, y), the multi-spectral ($\lambda$), polarization (p), and time (t).

    \begin{figure}[h]
    \centering
    \includegraphics[width=0.5\textwidth]{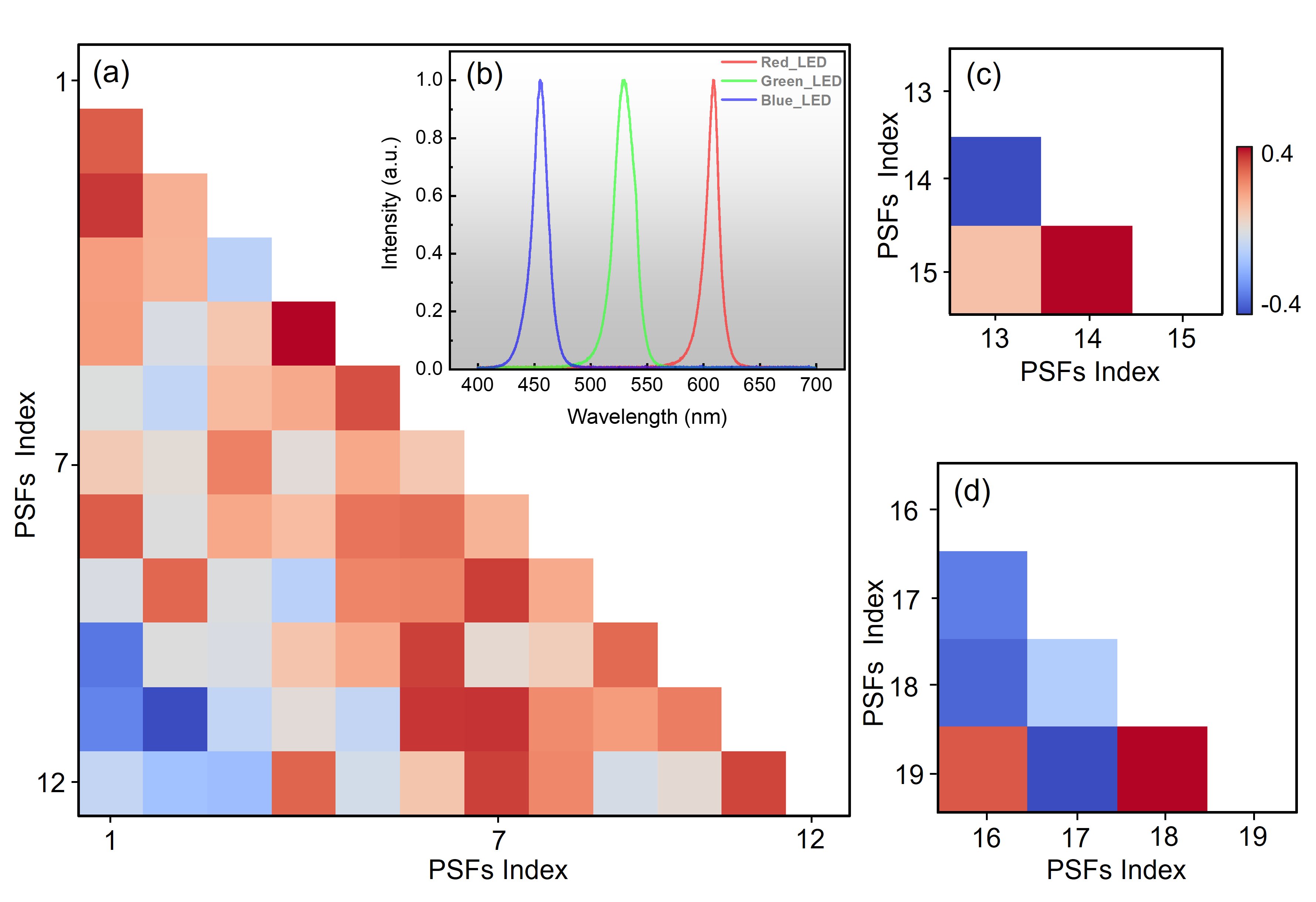}
    \caption{The correlation analysis of pre-calibration PSFs within different PSF sets. Pearson correlations of (a) the subset of the 12 spectral-polarization joint encoded PSFs (Set \#1: corresponding to the PSFs index from 1 to 12), (c) the subset of the polarization-independent multi-spectral (RGB) PSFs (Set \#3: corresponding to the PSFs index from 13 to 15), and (d) the linear polarization encoded PSFs under cascading white light illumination (Set \#2: corresponding to the PSFs index from 16 to 19). The values of Pearson correlation are all within a range from -0.4 to 0.4. (b) The spectra of the input three monochromatic ($\sim$10 nm) LEDs (Red, Green, and Blue). }
    \label{fig:2}  
    \end{figure}

    Since the wavelength and polarization of the real scene are continuously varying, j in Equation 1 should also be mathematically infinite. As a proof-of-concept demonstration, here we only calibrate three different colors (red, green, and blue), with a white light source that is combined by three monochromatic LEDs (see Supporting Information S1). All measurements in our experiment are all under the illumination of this white light source. Thus, the experimental raw data is the sum of different convolutions of the target object and different PSFs that contain only the three colors. We can also transform Equation 1 into the form of matrix multiplication (see the Supporting Information S2):
    \begin{equation}
        Y = AX,
    \end{equation}   
    where \emph{Y} is the column vector from raw data, \emph{X} is the column vector of the multi-dimensional object, and \emph{A} is the calibration matrix, which is juxtaposed and reshaped by the corresponding modalities’ PSFs. In order to accommodate different tasks with various imaging dimensions, it is possible to select and combine only a subset of the pre-calibrated PSFs to form the matrix A, rather than utilizing all of the PSFs to generate a larger matrix. This approach effectively reduces the matrix size and computational complexity, while also reducing computational resources in accordance with the actual modality requirements (see Supporting Information S3).

    We calculated the Pearson correlation of the sunsets of the PSFs corresponding to different dimension modalities, as shown in Figure 2a, c, and d. The absolute value of Pearson's coefficient between any two PSFs is less than 0.4, which can indicate a weak correlation between the PSFs. Thus, we can independently reconstruct the results of the switchable dimensional modalities. Figure 2b shows the spectra of the 3 monochromatic LED sources employed in the experiment, with the full width at half-maximum (FWHM) about 10 nm. By choosing the corresponding PSFs and under-sampling the different super-rows, we can reconstruct the images in the corresponding dimensional modality using the CS algorithm\cite{donoho2006compressed,liu2021self,wang2023dual,song2024computational} and the two-step iterative shrinkage/thresholding algorithm (TwIST)\cite{bioucas2007new}:
    \begin{equation}
    \hat{X} = arg\min_{X \geq 0} {\lVert y - AX \rVert}_2^2 + \tau {\lVert \Psi X \rVert}_1,
    \label{eq:inverse solve}
    \end{equation}
    where $\Psi$ is an exchange matrix that maps the object vector \emph{X} to a sparse representation, and $\tau$ is a tunable equilibrium parameter.

    \subsection{Results of Diffuser-mCam in temporal-compressive imaging modality}
    \begin{figure}[h]
    \centering
    \includegraphics[width=0.5\textwidth]{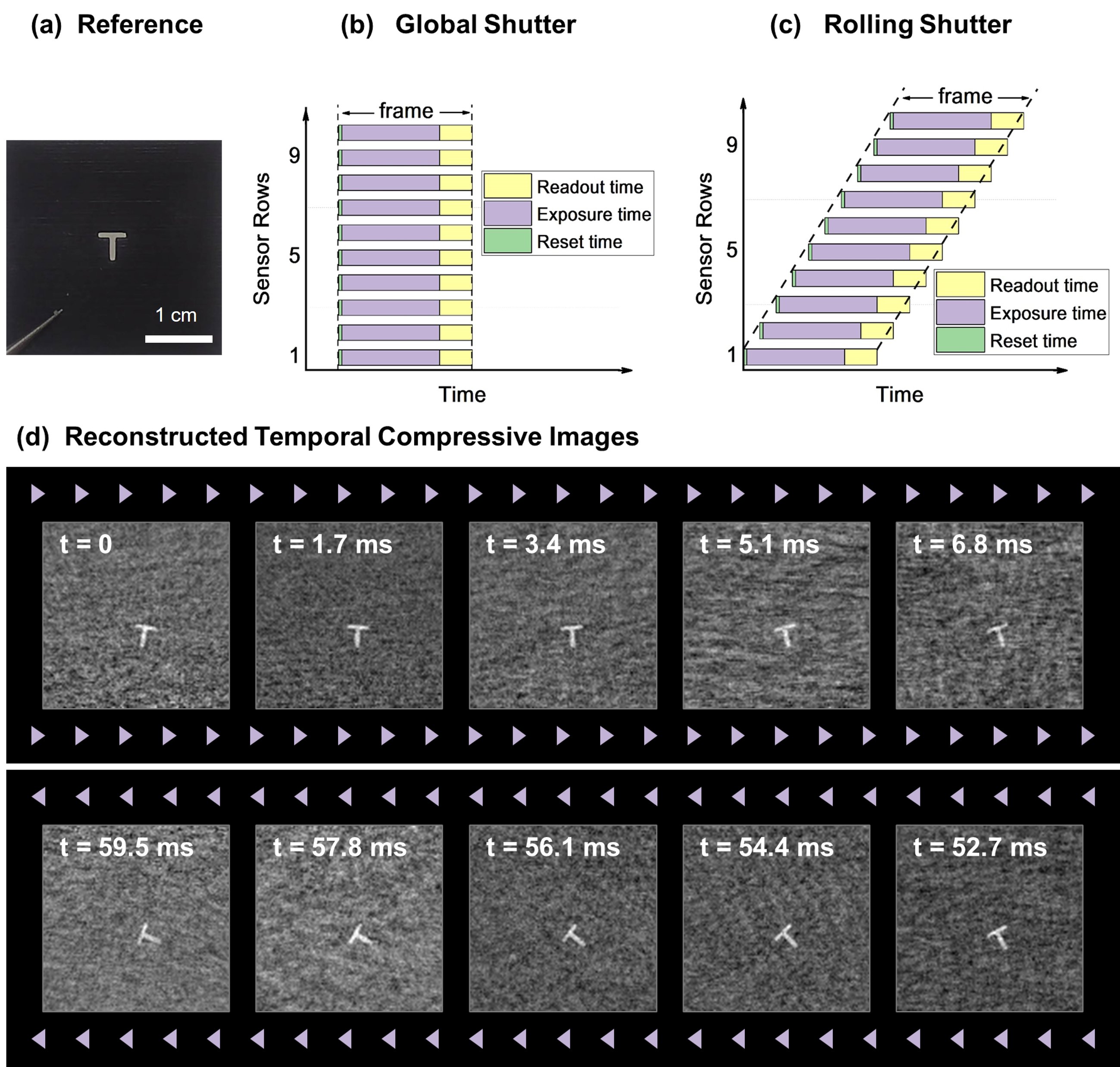}
    \caption{Reconstructed snapshot temporal compressive imaging results (108×132 pixels) of a dynamic object enabled by Diffuser-mCam in the temporal modality. (a) The reference image of the transmission object of the letter “T”, which is fixed on the rotating stage at an angular velocity of 1800 degrees per second. (b) Diagram of the global Shutter exposure mechanism, in which each row of the CMOS sensor begins and ends the exposure period simultaneously. (c) Diagram of the rolling Shutter exposure mechanism, in which each row of the CMOS sensor starts its exposure at a different time. The dynamic information of temporal objects is encoded into different rows of the raw data (i.e., super-rows) in rolling shutter mode. (d) The 36-frame temporal compressive images at t = 0, 1.7, 3.4, …, 57.8, 59.5 ms (corresponding to frame \#1, \#2, \#3, …,  \#35, \#36), which were reconstructed from a single-shot raw image.}
    \label{fig:3}  
    \end{figure}

    Figure 3 presents the reconstructed temporal compressive results (108×132 pixels) of a dynamic scene, with the sampling rate of 35.1 \% per channel, enabled by the single-shot of Diffuser-mCam in its temporal modality. In this case, the calibration matrix A in Equations 2 \& 3 is only related to a single PSF, noted as white-natural, without the recognition of spectral and polarization, corresponding to the PSFs' index of 20. It is unnecessary to reconstruct results such as spectral and polarization imaging, as the task-oriented approach allows for the efficient reduction of computational resource requirements. The temporal information of the dynamic scene is encoded into the super-rows due to the rolling-shutter exposure mode (Figure 1). In contrast to the conventional global shutter mode (Figure 3b), the rolling shutter mode (Figure 3c) initiates exposure sequentially across sensor rows, introducing a time delay between adjacent rows. We divided the raw data into 36 super-rows to balance the temporal and image resolution of the reconstructed results. A 3D-printed transmission object with the letter “T” (Figure 3a) is mounted on a rotation stage as the dynamic object in the experiment. The clockwise rotation speed is set to 1800 °/s. The original max frame rate of the CMOS sensor is 14 frames per second (fps). By using each super-row of the single-shot raw image as the input data, we reconstruct the 36-frame temporal compressive images (see Supporting Information S4). Figure 3d shows the selected example of the temporal compressive images, which correspond to a frame rate of 588 fps. Thus, Diffuser-mCam supports a 42-fold improvement in imaging frame rate, compared with the frame rate of the employed CMOS sensor (14 Hz).

    \subsection{Results of Diffuser-mCam in multi-spectral imaging modality}
    \begin{figure*}[htbp]
    \centering
    \includegraphics[width=0.8\textwidth]{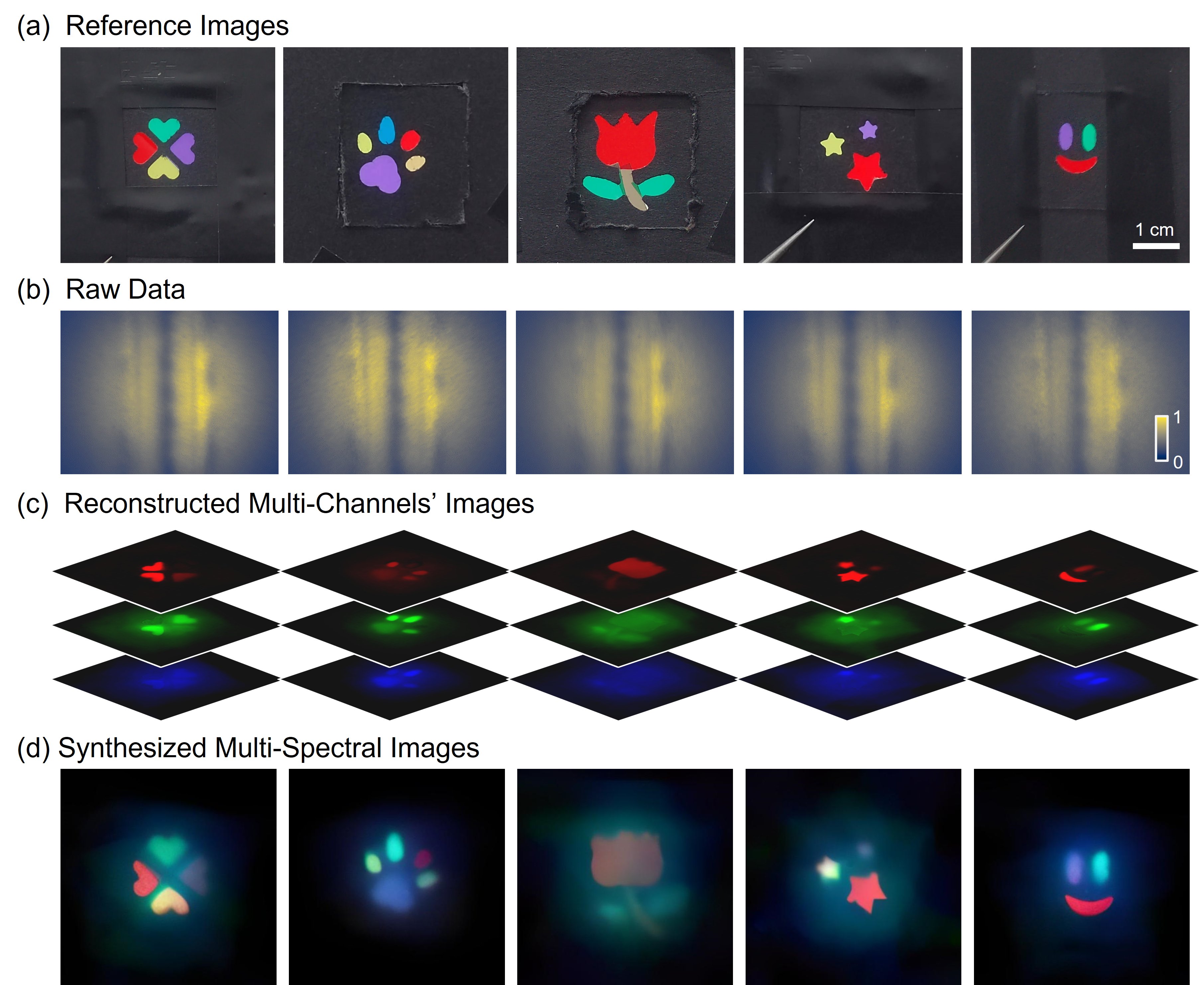}
    \caption{Experimental results of the static color objects, which are reconstructed from single-shot raw images by Diffuser-mCam in its multi-spectral imaging modality. (a) Reference images of the “clover”, “claw”, “flower”, “stars”, and “face” objects. These objects are made by pasting different colored transparent plastic sheets on the corresponding hollowed-out objects. (b) The corresponding raw images of the objects, which are directly captured by Diffuser-mCam. (c) Reconstructed multi-spectral images of the objects in separated channels (128×128 pixels), by using different spectral PSFs (red, green, and blue). (d) Synthesized multi-spectral RGB images with the inputs of (c).}
    \label{fig:4}  
    \end{figure*}    

    Diffuser-mCam adapts well to both dynamic and static scenes. If the scene exhibits slowly changing or quasi-static conditions in comparison to the sensor's frame rate, each super-row is able to be expanded, even to encompass the entire raw data. Figure 4 reports the reconstructed results of Diffuser-mCam towards the static colorful objects in multi-spectral imaging modality. In this case, the calibration matrix A is related to the set containing polarization-independent multi-spectral PSFs, noted as red-natural, green-natural, and blue-natural, corresponding to PSFs index from 13 to 15. The five static colorful objects used in the experiment are shown in Figure 4a, which resemble the shapes of “clover”, “claw”, “flower”, “stars”, and “face”. These objects were created by cutting out the corresponding shapes from commercial common black cardboard and sticking on transparent PVC colored filters. Figure 4(b) presents the images of the effective raw data of each object. As shown in Figure 4c, the output images in the red, green, and blue channels are reconstructed in a single step using the compressed sensing algorithm, respectively. The resolution of these reconstructed RGB images is 128×128 pixels. To make a fairer comparison of dynamic reconstruction performance, we under-sampled the region to the size of a single super-row (10×500 pixels, denoted as Y in Equation 3) and kept the sampling rate per channel the same as in dynamic reconstruction (10.2 \%, see Supporting Information S3). However, this setup would degrade the quality of the reconstructed images in Figure 4c\&d due to the low sampling rate. Figure 4d presents the synthesized multispectral RGB images. The completion of the five sets of multispectral modal imaging experiments for the five different objects is accomplished by a simple alteration of the different objects. The reconstructed multi-dimensional (RGB) images and spectral (see Supporting Information S5) illustrated the strong capacity of Diffuser-mCam to recover the light field in multi-spectral modality.

    \subsection{Results of Diffuser-mCam in polarization-Stocks imaging modality}
    \begin{figure*}[htbp]
    \centering
    \includegraphics[width=0.8\textwidth]{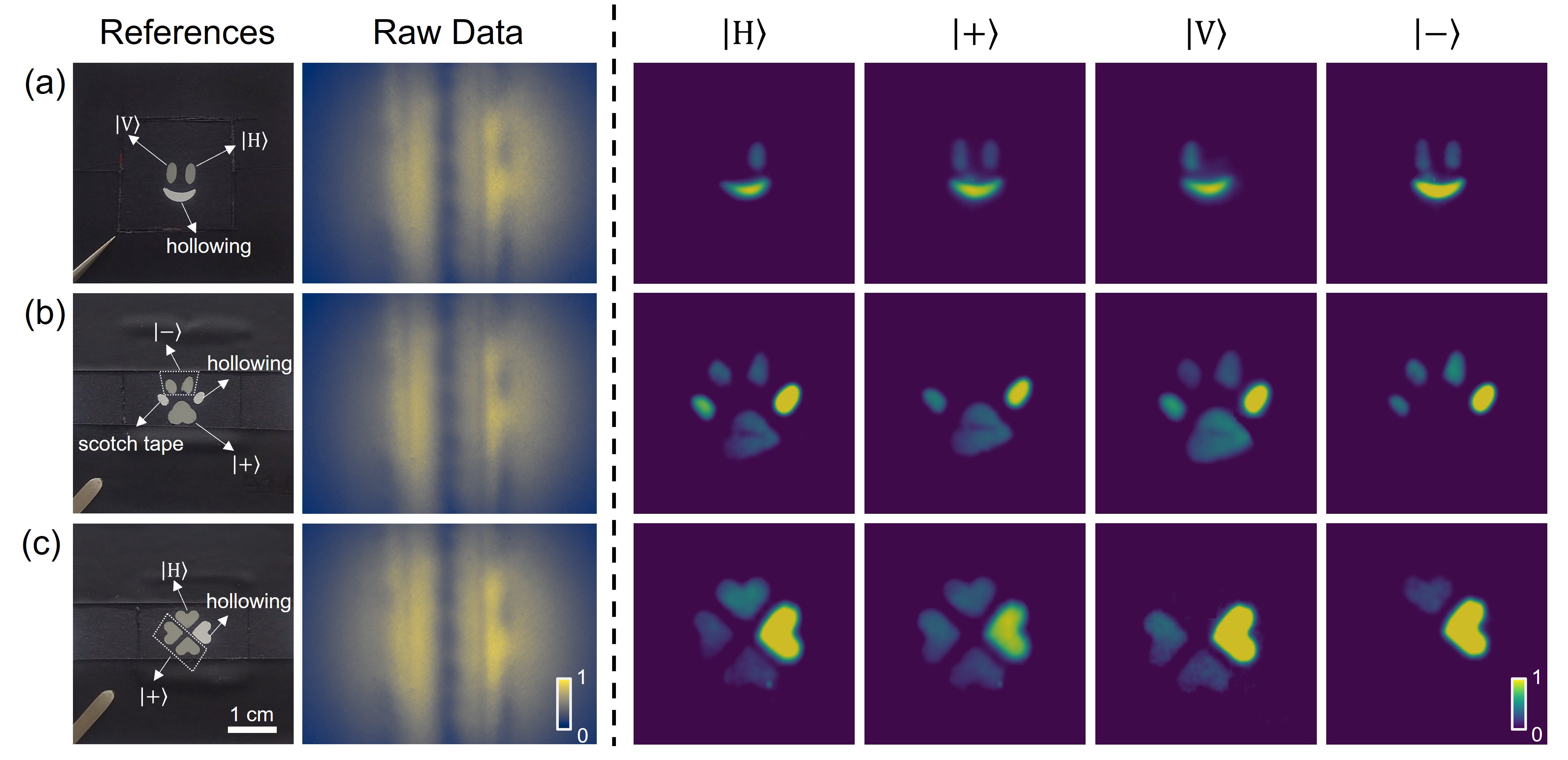}
    \caption{Experimental results of Diffuser-mCam in the polarization imaging modality. The reference images of the used static objects (a) “face”, (b) “claw”, and (c) “clover”, which were made by carefully pasting thin-film polarizers of different linear directions on the different transmission parts, are shown in the left column. The corresponding raw data is adjacent in the right column of them. The corresponding reconstructed images of the four polarization channels: |H>, |+>, |V>, |->  are shown on the right side of the dashed line. All the reconstructed image resolutions are 128×128 pixels. }
    \label{fig:5}  
    \end{figure*}  

     For static polarization objects, Figure 5 reports the reconstructed results of Diffuser-mCam in its polarization imaging modality. The calibration matrix A is supposed to be related to the set that contains the PSFs of white light with four linear polarizations, noted as white-$\vert H\rangle$, white-$\vert +\rangle$, white-$\vert V\rangle$, and white-$\vert -\rangle$ (corresponding to PSFs index from 16 to 19). The used transmission objects of “face”, “claw”, and “clover”, are shown in the first column in Figure 5. To design the polarization objects, the regions of each object are pasted with line polarizers with varied polarization directions, some areas, denoted ‘hollowing’ in Figure 5(a), are covered with stretched transparent scotch tape. The internal molecular structure of transparent scotch tape undergoes modification under the influence of applied tensile stress, resulting in the emergence of polarization properties, known as the “photoelastic effect” phenomenon. The resolution of these reconstructed polarization images is 128×128 pixels. By under-sampling the region to the size of a single super-row (10×500 pixels, denoted as Y in Equation 3) and keeping the sampling rate per channel the same as in dynamic reconstruction (7.6 \%, see Supporting Information S3), the corresponding polarization images under the four line polarization channels can be reconstructed all at once through the CS algorithm, as shown in the right four columns of Figure 5. As illustrated in Figure 5, transparent objects with polarization characteristics are indistinguishable under conventional photography, such as the two “eyes” of the “face” (Figure 5a), the “fingers” and “palm” of the “claw” (Figure 5b), and the various “leaves” of the “clover” (Figure 5c). However, they are all successfully classified under different polarization channels. This finding demonstrates the potential of Diffuser-mCam for polarization detection and classification of objects, such as crystal birefringence characterization and biomedical imaging.

    The Stokes parameter is an effective means of describing the polarization state of light. Furthermore, in our linearly polarized scene, we calculated the Stokes images by:
    \begin{equation}
    \begin{aligned}
    S_0 &= \frac{I_{\vert H\rangle} + I_{\vert +\rangle} + I_{\vert V\rangle} + I_{\vert -\rangle}}{4}, \\
    S_1 &= \frac{I_{\vert H\rangle} - I_{\vert V\rangle}}{2}, \\
    S_2 &= \frac{I_{\vert +\rangle} - I_{\vert -\rangle}}{2},
    \end{aligned}
    \end{equation}
    where $I_{\vert H\rangle}$ , $I_{\vert +\rangle}$ , $I_{\vert V\rangle}$ and $I_{\vert _\rangle}$ are the reconstructed polarization images of different channels presented in Figure 5. S0 represents the total intensity of the polarized light, S1 represents the difference between the horizontally (i.e. $\vert H\rangle$ ) and vertically (i.e. $\vert V\rangle$ ) polarized components, and S2 represents the difference in the intensity of the line polarization in the ±45° (i.e. $\vert +\rangle$ and  $\vert -\rangle$ ) direction. S3 is employed for the description of the circularly polarized properties of the light, and it is not computed in our experiments.

    We also calculate the images of the degree of linear polarization (DoLP) and the angle of linear polarization (AoLP), based on the Stokes parameter calculated in Equation 4. DoLP represents the proportion of total light intensity accounted for by the linear polarization light. The closer DoLP’s value is to 1, the stronger the linear polarization of total light is. AoLP represents the angle between the direction of polarization light and the horizontal direction, i.e., the vibration direction of linear polarization light.
    \begin{equation}
    \begin{aligned}
    \text{DoLP} &= \frac{\sqrt{S_1^2 + S_2^2}}{S_0}, \\
    \text{AoLP} &= \frac{1}{2} \arctan\left(\frac{S_2}{S_1}\right).
    \end{aligned}
    \end{equation}

    \begin{figure*}[htbp]
    \centering
    \includegraphics[width=0.8\textwidth]{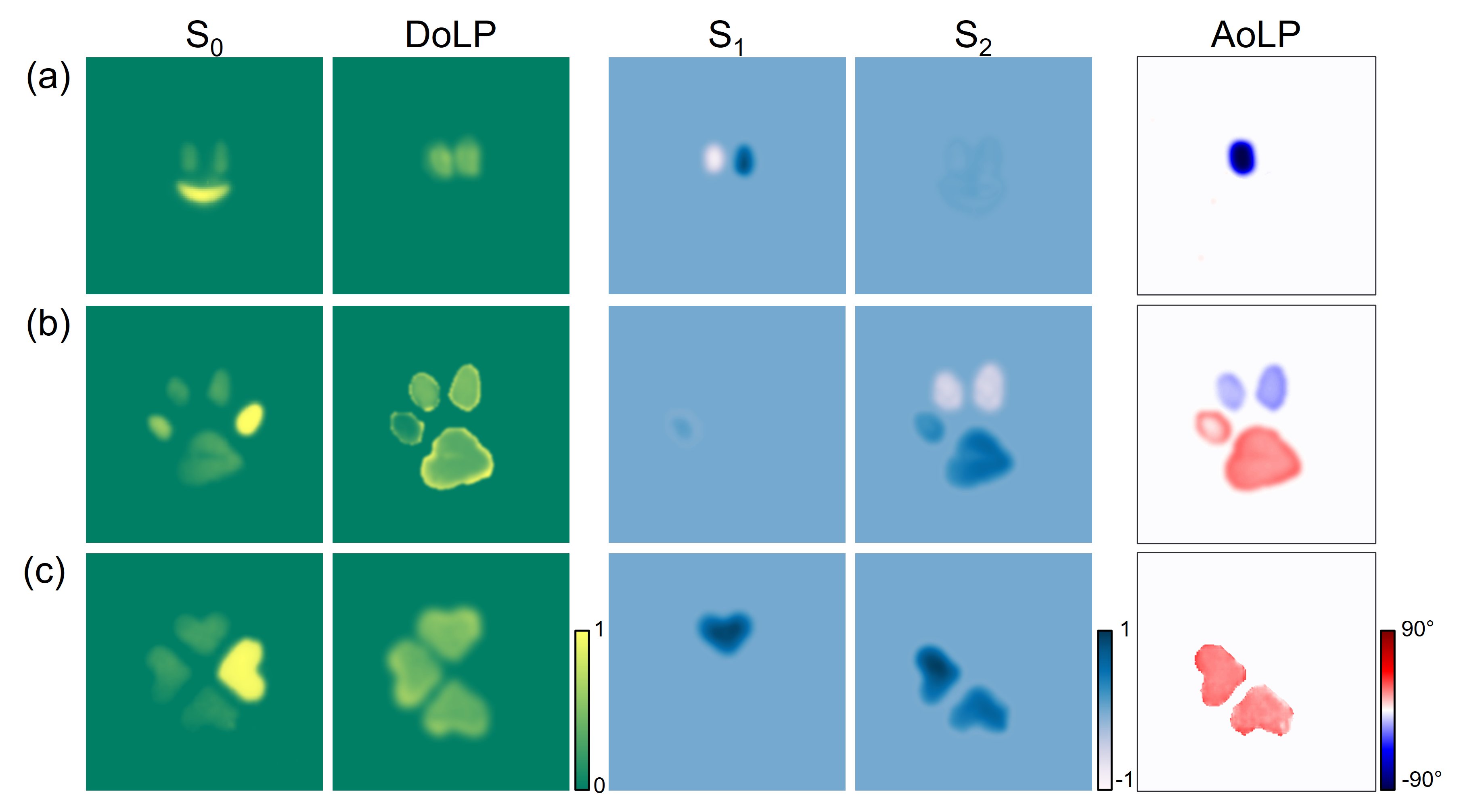}
    \caption{The Stocks Analysis images (S0, DoLP, S1, S2, AoLP) of the three see-through polarization objects: (a) “face”, (b) “claw”, (c) “clover”.}
    \label{fig:6}  
    \end{figure*} 

     Figure 6 presents the Stokes analysis images (S0, DoLP, S1, S2, AoLP) of the three polarization objects. A comparison of the DoLP image with the S0 image enables the recognition of the proportion of line polarized light in disparate regions of the object. For instance, as illustrated in Figures 5\&6(a), the "mouth" of the "face" is not polarized, and its intensity is measured at zero in the DoLP image. A comparison of the DoLP image with the AoLP image enables the discernment of the linear polarization angles of the distinct polarized regions, thereby supplying a reference for the analysis of the target transparent objects' properties. As illustrated in Figure 6(c), the three "leaves" of the line polarizer affixed to the "clover" are indistinguishable from one another in the directly captured image (reference image), S0, and DoLP image. The discernible difference in polarization angle was only evident in the AoLP image. By further analyzing the imaging results of polarization-Stokes modality of the same transparent object, the polarization characteristics of each object can be further classified and identified for tasks that are indistinguishable to the naked eye, such as camouflage identification and stress detection. 
    
    While the demonstration has been conducted to illustrate the imaging capabilities of Diffuser-mCam in specific spectral-only and polarization-only modalities using a static scene as an example, it is imperative to emphasize that the results for the remaining moments of these two modalities can be reconstructed by sampling other super rows.

    \subsection{Reconstructed results of Diffuser-mCam in the five-dimensional modality}
    \begin{figure*}[htbp]
    \centering
    \includegraphics[width=\textwidth]{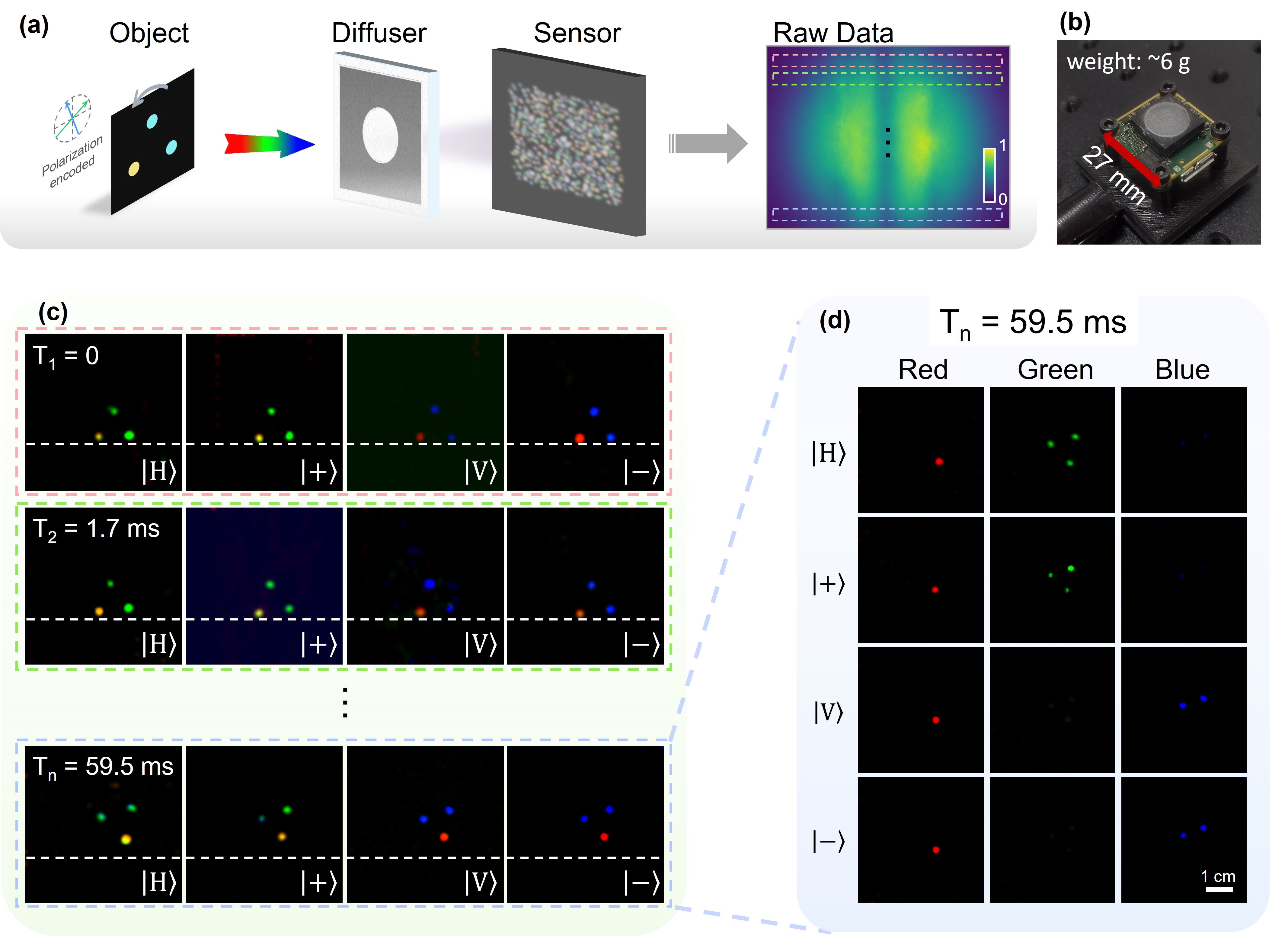}
    \caption{The full 5D modality imaging and classification experimental results enabled by Diffuser-mCam. (a) The dynamic spectral-polarization encoded scene (three-points, each point is encoded with different spectral and polarization information) is captured by the Diffuser-mCam and read out as speckle-like raw data. (b) Photograph of the Diffuser-mCam (with a size of 27 mm × 27 mm × 7 mm, and a weight of 6g). (c) The temporal compressive multi-spectral (RGB) imaging results of 4 polarization channels at different time points (0, 1.7 ms, …, 59.5 ms), which are reconstructed from the corresponding colored dashed super-rows. The auxiliary white dashed lines are used to help identify the slight rotation of objects between two adjacent frames.  (d) The full spectral-polarization channels’ (3 × 4) results at Tn = 59.5 ms.}
    \label{fig:6}  
    \end{figure*} 

    In this section, we demonstrate the proof-of-principle for the reconstruction of Diffuser-mCam in the 5D imaging modality. The photograph of Diffuser-mCam, with a size of 27 mm × 27 mm × 7 mm and a weight of ~6g, is presented in Figure 7b. We employed the set of the full 12 spectral-polarization joint encoded PSFs (corresponding to PSFs index from 1 to 12) to compose the calibration matrix A. In this case, the maximum sampling rate is 2.5 \% per channel (see Supporting Information S3). The employed transmissive object, as presented in Figure 7a, is composed of three small round points, which are designed to satisfy the sparsity assumption of the CS algorithm\cite{kabuli2025designinglenslessimagingsystems}.  Each round point has a diameter of 1.5 mm. The one point is affixed with a yellow PVC sheet and a ($\vert H\rangle$+$\vert +\rangle$)/2 linear polariser. The other two points are affixed with a cyan PVC sheet and a ($\vert V\rangle$+$\vert -\rangle$)/2 linear polariser. The object was pasted on a counterclockwise rotating stage during the measurement, with a speed of 1800°/s. In the course of our experiments, we illuminated the dynamic, spectral polarization object with white light. The transmitted light was modulated by the diffuser and the polarized mask. The CMOS sensor snapshots the speckle-like raw data in rolling shutter mode. By under-sampling the corresponding super-rows, the temporal compressive spectral-polarization images are reconstructed through the CS algorithm, as shown in Figure 7c (the three reconstructed images under the RGB channels have been synthesized into a single RGB color image for better presentation). The 12 spectral-polarization images (3 spectral channels × 4 polarization channels) in the same time channel can be reconstructed simultaneously. The results for the remaining moments must be recovered by under-sampling the corresponding super-rows data once more and recomputing the inverse problem (Equation 3). Figure 7d presents the 12 spectral-polarization images at Tn. The minimum time resolution of Diffuser-mCam is 1.7 ms, and the maximum frame rate is 588 fps. In other words, Diffuser-mCam is able to encode 12 (spectral-polarization channels) × 36 (time channels) = 432 channels (i.e., 6048 channels per second) into a snapshot of raw data and effectively reconstruct the corresponding results. 

\section{Discussion and Conclusion} \label{sec:dac}
    %\lipsum[6-7]
     In summary, we have proposed and demonstrated the compact snapshot multi-dimensional imaging and classification diffuser camera (Diffuser-mCam). The Diffuser-mCam is a compact, low-cost, lightweight imaging system with switchable imaging modalities for different sensing scenarios, making it ideal for deployment in space-constrained practical applications. In our experiments, we enhance the encoding of spectral and polarization information of the incident light field by scattering medium and polarization mask, and we also take advantage of the roll-up shutter effect of the CMOS camera to record the dynamic time information of the incident light field in different rows of the raw data, and finally realize the encoding of 432 channels of image information in a single snapshot of grayscale raw data (i.e. 6048 channels per second). After a straightforward pre-calibration of the sets of PSFs of all possible optical modalities, Diffuser-mCam is able to freely switch imaging dimensional modalities without any experimental modifications. We can directionally selectively recover the results corresponding to different dimensional modalities (or all five-dimensional modalities) from the same raw data (as demonstrated in Figure 3-7). Diffuser-mCam realizes a resource-efficient computational multi-dimensional imaging paradigm in accordance with real-world task requirements. This on-demand framework can be oriented to the actual task requirements to reconstruct the results of the specific modality without having to reconstruct the full five-dimensional results every time. 

    The performance of the Diffuser-mCam is largely limited by the device performance versus the number of pre-calibrated PSFs. Theoretically, if additional imaging dimensional modalities (e.g., depth, angular momentum, etc.) can be encoded and pre-calibrated\cite{boominathan2021recent}, the imaging dimensions of the Diffuser-mCam can be further enhanced. Similarly, if more channels can be calibrated in the same dimensional modality, e.g., boosting the number of multispectral channels under natural light illumination to hyperspectral levels (channels >100), the resolution for each modality can be further improved. This requires a better method for light field modulation to enhance the multidimensional encoding capabilities\cite{zhang2024robust,wang2024multi,tian2024miniaturized}. The performance of the COMS sensor, in particular the frame rate, bandwidth, and offline delay time of the rolling shutter mode, are the few parameters that restrict the temporal resolution of the Diffuser-mCam. The high-performance sCMOS or EMCCD offers a solution to this problem\cite{antipa2019video,weinberg2020100}, but their utilization will increase the cost and the size. Optimization of algorithms would also improve its performance\cite{gao2024motion}, e.g., the use of deep learning\cite{cao2024neural} or quantum machine learning\cite{xiao2024quantum} can further increase the resolution and number of channels of the Diffuser-mCam.
\section{Experimental Section} \label{sec:Exp}
Experimental Setup: Diffuser-mCam is composed of a CMOS sensor (daA2500-14um, Basler), a thick diffuser (DW105-120, LBTEK), and a self-made polarized mask. The polarized mask, which is pasted close to the sensor plane, comprises four vertical strips of thin-film polarization filters, each exhibiting a distinct direction of linear polarization (${\vert H\rangle}$ , ${\vert+\rangle}$ , ${\vert V\rangle}$ and ${\vert-\rangle}$), arranged in parallel to the sensor’s column. The four vertical polarization filters were obtained by meticulous cutting of a large thin film polarizer (FLP-VIS-100, LBTEK). The Diffuser was mounted on a 3D-printed stand, as close as possible to the sensor and the polarization mask. The stand was carefully pasted onto the bare sensor chip with precision to ensure the stability of the Diffuser-mCam apparatus employed in the experiments. (as shown in Figure 7b). The Diffuser-mCam was fixated and illuminated with the collimated incident white light source. The incident white light in our experiment is cascaded by three monochromatic red, green, and blue incoherent LEDs (GCI-060401, GCI-060403, GCI-060404, DHC). During the measurement phase, the various static multidimensional transmissive objects referenced in the main text were positioned in the light path at a distance of 50 cm from the Diffuser-mCam. During the measurement, we set 5×5 pixels of the CMOS sensor as 1 super pixel to balance the resolution and the field of view (FOV), and the resolution of CMOS sensor is 388×518 pixels. The middle region of 360×500 pixels are designated as the effective raw data. The exposure time of the CMOS sensor is set to 4000 $\mu$s, and the snapshot row data is captured in rolling-shutter mode. The dynamic objects were affixed in the rotation mount (DDR25, Thorlabs) at a rotation velocity of 1800°/s, without changing the other experimental settings. The spectra in this work are measured by the commercial spectrometer (USB4000-UV-VIS-ES, Ocean Optics).

PSFs Pre-Calibration: The Diffuser-mCam was fixated, and four sets of the PSFs corresponding to different imaging modalities should be pre-calibrated according to the forward imaging model. For pre-calibration, we put a 75 $\mu$m pinhole in the center of the object plane, instead of the above multi-dimensional objects, at a distance of 50 cm from the Diffuser-mCam. 1) To reconstruct the results of the full five-dimensions modality, the set of 12 spectral-polarization joint modulated PSFs should be generated, noted as red-${\vert H\rangle}$, red-${\vert+\rangle}$, red-${\vert V\rangle}$, red-${\vert-\rangle}$, green-${\vert H\rangle}$, green-${\vert+\rangle}$, green-${\vert V\rangle}$, green-${\vert-\rangle}$, blue-${\vert H\rangle}$, blue-${\vert+\rangle}$, blue-${\vert V\rangle}$, blue-${\vert-\rangle}$ (corresponding to the PSFs index from 1 to 12). An auxiliary angle-adjustable polarizer (OPPF1-VIS, JCOPTIX) is employed to modulate the incident monochromatic LED light, thereby generating linear polarization light at four directions (${\vert H\rangle}$ , ${\vert+\rangle}$ , ${\vert V\rangle}$ and ${\vert-\rangle}$). By substituting the red, green, and blue LED light sources, this set of PSFs can be pre-calibrated. 2) To reconstruct the results of the multi-spectral (RGB) modality, the set of 3 polarization-independent multi-spectral PSFs should be generated, noted as red-natural, green-natural, and blue-natural (corresponding to the PSFs index from 13 to 15). By only substituting the red, green, and blue LED light sources without the auxiliary polarizer, this set of PSFs can be pre-calibrated. 3) To reconstruct the results of the polarization-Stocks modality, the set of 4 spectral-independent polarization PSFs should be generated, noted as white-${\vert H\rangle}$, white-${\vert+\rangle}$, white-${\vert V\rangle}$, and white-${\vert-\rangle}$ (corresponding to PSFs index from 16 to 19). By only substituting the linear angle of the auxiliary polarizer with the cascaded white light, this set of PSFs can be pre-calibrated. 4) To reconstruct the results of the temporal modality, a polarization spectral-independent PSF should be generated (corresponding to the PSFs index of 20). (see Supporting Information S3). The PSFs can be pre-calibrated using the cascaded white light without the auxiliary polarizer.
\section*{Acknowledgements} \label{sec:acknowledgements}
    %\lipsum[8]
    This work was supported by the National Natural Science Foundation of China (U23A20481, 62275010) and the Fundamental Research Funds for the Central Universities (KG16-3549-01). 
\appendix*
\section{Supporting Information} \label{sec:appendix}
    %\lipsum[9-11]
    The appendix and Supporting Information that support the results presented in this study are available from the corresponding author upon reasonable request.

\bibliographystyle{apsrev4-1}  % 选择样式（如plain、apsrev4-1等）
\bibliography{references}  % 引用BibTeX文件（不含.bib后缀）

\begin{thebibliography}{55}%
\makeatletter
\providecommand \@ifxundefined [1]{%
 \@ifx{#1\undefined}
}%
\providecommand \@ifnum [1]{%
 \ifnum #1\expandafter \@firstoftwo
 \else \expandafter \@secondoftwo
 \fi
}%
\providecommand \@ifx [1]{%
 \ifx #1\expandafter \@firstoftwo
 \else \expandafter \@secondoftwo
 \fi
}%
\providecommand \natexlab [1]{#1}%
\providecommand \enquote  [1]{``#1''}%
\providecommand \bibnamefont  [1]{#1}%
\providecommand \bibfnamefont [1]{#1}%
\providecommand \citenamefont [1]{#1}%
\providecommand \href@noop [0]{\@secondoftwo}%
\providecommand \href [0]{\begingroup \@sanitize@url \@href}%
\providecommand \@href[1]{\@@startlink{#1}\@@href}%
\providecommand \@@href[1]{\endgroup#1\@@endlink}%
\providecommand \@sanitize@url [0]{\catcode `\\12\catcode `\$12\catcode `\&12\catcode `\#12\catcode `\^12\catcode `\_12\catcode `\%12\relax}%
\providecommand \@@startlink[1]{}%
\providecommand \@@endlink[0]{}%
\providecommand \url  [0]{\begingroup\@sanitize@url \@url }%
\providecommand \@url [1]{\endgroup\@href {#1}{\urlprefix }}%
\providecommand \urlprefix  [0]{URL }%
\providecommand \Eprint [0]{\href }%
\providecommand \doibase [0]{http://dx.doi.org/}%
\providecommand \selectlanguage [0]{\@gobble}%
\providecommand \bibinfo  [0]{\@secondoftwo}%
\providecommand \bibfield  [0]{\@secondoftwo}%
\providecommand \translation [1]{[#1]}%
\providecommand \BibitemOpen [0]{}%
\providecommand \bibitemStop [0]{}%
\providecommand \bibitemNoStop [0]{.\EOS\space}%
\providecommand \EOS [0]{\spacefactor3000\relax}%
\providecommand \BibitemShut  [1]{\csname bibitem#1\endcsname}%
\let\auto@bib@innerbib\@empty
%</preamble>
\bibitem [{\citenamefont {Gao}\ \emph {et~al.}(2010)\citenamefont {Gao}, \citenamefont {Kester}, \citenamefont {Hagen},\ and\ \citenamefont {Tkaczyk}}]{gao2010snapshot}%
  \BibitemOpen
  \bibfield  {author} {\bibinfo {author} {\bibfnamefont {L.}~\bibnamefont {Gao}}, \bibinfo {author} {\bibfnamefont {R.~T.}\ \bibnamefont {Kester}}, \bibinfo {author} {\bibfnamefont {N.}~\bibnamefont {Hagen}}, \ and\ \bibinfo {author} {\bibfnamefont {T.~S.}\ \bibnamefont {Tkaczyk}},\ }\href@noop {} {\bibfield  {journal} {\bibinfo  {journal} {Optics Express}\ }\textbf {\bibinfo {volume} {18}},\ \bibinfo {pages} {14330} (\bibinfo {year} {2010})}\BibitemShut {NoStop}%
\bibitem [{\citenamefont {Arguello}\ and\ \citenamefont {Arce}(2012)}]{arguello2012rank}%
  \BibitemOpen
  \bibfield  {author} {\bibinfo {author} {\bibfnamefont {H.}~\bibnamefont {Arguello}}\ and\ \bibinfo {author} {\bibfnamefont {G.~R.}\ \bibnamefont {Arce}},\ }\href@noop {} {\bibfield  {journal} {\bibinfo  {journal} {IEEE Transactions on Image Processing}\ }\textbf {\bibinfo {volume} {22}},\ \bibinfo {pages} {941} (\bibinfo {year} {2012})}\BibitemShut {NoStop}%
\bibitem [{\citenamefont {Nakagawa}\ \emph {et~al.}(2014)\citenamefont {Nakagawa}, \citenamefont {Iwasaki}, \citenamefont {Oishi}, \citenamefont {Horisaki}, \citenamefont {Tsukamoto}, \citenamefont {Nakamura}, \citenamefont {Hirosawa}, \citenamefont {Liao}, \citenamefont {Ushida}, \citenamefont {Goda} \emph {et~al.}}]{nakagawa2014sequentially}%
  \BibitemOpen
  \bibfield  {author} {\bibinfo {author} {\bibfnamefont {K.}~\bibnamefont {Nakagawa}}, \bibinfo {author} {\bibfnamefont {A.}~\bibnamefont {Iwasaki}}, \bibinfo {author} {\bibfnamefont {Y.}~\bibnamefont {Oishi}}, \bibinfo {author} {\bibfnamefont {R.}~\bibnamefont {Horisaki}}, \bibinfo {author} {\bibfnamefont {A.}~\bibnamefont {Tsukamoto}}, \bibinfo {author} {\bibfnamefont {A.}~\bibnamefont {Nakamura}}, \bibinfo {author} {\bibfnamefont {K.}~\bibnamefont {Hirosawa}}, \bibinfo {author} {\bibfnamefont {H.}~\bibnamefont {Liao}}, \bibinfo {author} {\bibfnamefont {T.}~\bibnamefont {Ushida}}, \bibinfo {author} {\bibfnamefont {K.}~\bibnamefont {Goda}},  \emph {et~al.},\ }\href@noop {} {\bibfield  {journal} {\bibinfo  {journal} {Nature Photonics}\ }\textbf {\bibinfo {volume} {8}},\ \bibinfo {pages} {695} (\bibinfo {year} {2014})}\BibitemShut {NoStop}%
\bibitem [{\citenamefont {Wang}\ \emph {et~al.}(2025)\citenamefont {Wang}, \citenamefont {Peng}, \citenamefont {Fang},\ and\ \citenamefont {Gao}}]{wang2025computational}%
  \BibitemOpen
  \bibfield  {author} {\bibinfo {author} {\bibfnamefont {Z.}~\bibnamefont {Wang}}, \bibinfo {author} {\bibfnamefont {Y.}~\bibnamefont {Peng}}, \bibinfo {author} {\bibfnamefont {L.}~\bibnamefont {Fang}}, \ and\ \bibinfo {author} {\bibfnamefont {L.}~\bibnamefont {Gao}},\ }\href@noop {} {\bibfield  {journal} {\bibinfo  {journal} {Optica}\ }\textbf {\bibinfo {volume} {12}},\ \bibinfo {pages} {113} (\bibinfo {year} {2025})}\BibitemShut {NoStop}%
\bibitem [{\citenamefont {Hua}\ \emph {et~al.}(2022)\citenamefont {Hua}, \citenamefont {Wang}, \citenamefont {Wang}, \citenamefont {Zou}, \citenamefont {Zhou}, \citenamefont {Li}, \citenamefont {Yan}, \citenamefont {Cao}, \citenamefont {Xiao}, \citenamefont {Tsai} \emph {et~al.}}]{hua2022ultra}%
  \BibitemOpen
  \bibfield  {author} {\bibinfo {author} {\bibfnamefont {X.}~\bibnamefont {Hua}}, \bibinfo {author} {\bibfnamefont {Y.}~\bibnamefont {Wang}}, \bibinfo {author} {\bibfnamefont {S.}~\bibnamefont {Wang}}, \bibinfo {author} {\bibfnamefont {X.}~\bibnamefont {Zou}}, \bibinfo {author} {\bibfnamefont {Y.}~\bibnamefont {Zhou}}, \bibinfo {author} {\bibfnamefont {L.}~\bibnamefont {Li}}, \bibinfo {author} {\bibfnamefont {F.}~\bibnamefont {Yan}}, \bibinfo {author} {\bibfnamefont {X.}~\bibnamefont {Cao}}, \bibinfo {author} {\bibfnamefont {S.}~\bibnamefont {Xiao}}, \bibinfo {author} {\bibfnamefont {D.~P.}\ \bibnamefont {Tsai}},  \emph {et~al.},\ }\href@noop {} {\bibfield  {journal} {\bibinfo  {journal} {Nature Communications}\ }\textbf {\bibinfo {volume} {13}},\ \bibinfo {pages} {2732} (\bibinfo {year} {2022})}\BibitemShut {NoStop}%
\bibitem [{\citenamefont {Lin}\ \emph {et~al.}(2022)\citenamefont {Lin}, \citenamefont {Pestourie}, \citenamefont {Roques-Carmes}, \citenamefont {Li}, \citenamefont {Capasso}, \citenamefont {Solja{\v{c}}i{\'c}},\ and\ \citenamefont {Johnson}}]{lin2022end}%
  \BibitemOpen
  \bibfield  {author} {\bibinfo {author} {\bibfnamefont {Z.}~\bibnamefont {Lin}}, \bibinfo {author} {\bibfnamefont {R.}~\bibnamefont {Pestourie}}, \bibinfo {author} {\bibfnamefont {C.}~\bibnamefont {Roques-Carmes}}, \bibinfo {author} {\bibfnamefont {Z.}~\bibnamefont {Li}}, \bibinfo {author} {\bibfnamefont {F.}~\bibnamefont {Capasso}}, \bibinfo {author} {\bibfnamefont {M.}~\bibnamefont {Solja{\v{c}}i{\'c}}}, \ and\ \bibinfo {author} {\bibfnamefont {S.~G.}\ \bibnamefont {Johnson}},\ }\href@noop {} {\bibfield  {journal} {\bibinfo  {journal} {Optics Express}\ }\textbf {\bibinfo {volume} {30}},\ \bibinfo {pages} {28358} (\bibinfo {year} {2022})}\BibitemShut {NoStop}%
\bibitem [{\citenamefont {Park}\ \emph {et~al.}(2020)\citenamefont {Park}, \citenamefont {Feng}, \citenamefont {Liang},\ and\ \citenamefont {Gao}}]{park2020snapshot}%
  \BibitemOpen
  \bibfield  {author} {\bibinfo {author} {\bibfnamefont {J.}~\bibnamefont {Park}}, \bibinfo {author} {\bibfnamefont {X.}~\bibnamefont {Feng}}, \bibinfo {author} {\bibfnamefont {R.}~\bibnamefont {Liang}}, \ and\ \bibinfo {author} {\bibfnamefont {L.}~\bibnamefont {Gao}},\ }\href@noop {} {\bibfield  {journal} {\bibinfo  {journal} {Nature Communications}\ }\textbf {\bibinfo {volume} {11}},\ \bibinfo {pages} {5602} (\bibinfo {year} {2020})}\BibitemShut {NoStop}%
\bibitem [{\citenamefont {Joshi}\ \emph {et~al.}(2025)\citenamefont {Joshi}, \citenamefont {Tiwari}, \citenamefont {Kahro}, \citenamefont {Xavier}, \citenamefont {Tahara}, \citenamefont {Kasikov}, \citenamefont {Kukli}, \citenamefont {Juodkazis}, \citenamefont {Tamm}, \citenamefont {Rosen} \emph {et~al.}}]{joshi2025interferenceless}%
  \BibitemOpen
  \bibfield  {author} {\bibinfo {author} {\bibfnamefont {N.}~\bibnamefont {Joshi}}, \bibinfo {author} {\bibfnamefont {V.}~\bibnamefont {Tiwari}}, \bibinfo {author} {\bibfnamefont {T.}~\bibnamefont {Kahro}}, \bibinfo {author} {\bibfnamefont {A.~P.~I.}\ \bibnamefont {Xavier}}, \bibinfo {author} {\bibfnamefont {T.}~\bibnamefont {Tahara}}, \bibinfo {author} {\bibfnamefont {A.}~\bibnamefont {Kasikov}}, \bibinfo {author} {\bibfnamefont {K.}~\bibnamefont {Kukli}}, \bibinfo {author} {\bibfnamefont {S.}~\bibnamefont {Juodkazis}}, \bibinfo {author} {\bibfnamefont {A.}~\bibnamefont {Tamm}}, \bibinfo {author} {\bibfnamefont {J.}~\bibnamefont {Rosen}},  \emph {et~al.},\ }\href@noop {} {\bibfield  {journal} {\bibinfo  {journal} {Journal of Physics: Photonics}\ } (\bibinfo {year} {2025})}\BibitemShut {NoStop}%
\bibitem [{\citenamefont {Yako}\ \emph {et~al.}(2023)\citenamefont {Yako}, \citenamefont {Yamaoka}, \citenamefont {Kiyohara}, \citenamefont {Hosokawa}, \citenamefont {Noda}, \citenamefont {Tack}, \citenamefont {Spooren}, \citenamefont {Hirasawa},\ and\ \citenamefont {Ishikawa}}]{yako2023video}%
  \BibitemOpen
  \bibfield  {author} {\bibinfo {author} {\bibfnamefont {M.}~\bibnamefont {Yako}}, \bibinfo {author} {\bibfnamefont {Y.}~\bibnamefont {Yamaoka}}, \bibinfo {author} {\bibfnamefont {T.}~\bibnamefont {Kiyohara}}, \bibinfo {author} {\bibfnamefont {C.}~\bibnamefont {Hosokawa}}, \bibinfo {author} {\bibfnamefont {A.}~\bibnamefont {Noda}}, \bibinfo {author} {\bibfnamefont {K.}~\bibnamefont {Tack}}, \bibinfo {author} {\bibfnamefont {N.}~\bibnamefont {Spooren}}, \bibinfo {author} {\bibfnamefont {T.}~\bibnamefont {Hirasawa}}, \ and\ \bibinfo {author} {\bibfnamefont {A.}~\bibnamefont {Ishikawa}},\ }\href@noop {} {\bibfield  {journal} {\bibinfo  {journal} {Nature Photonics}\ }\textbf {\bibinfo {volume} {17}},\ \bibinfo {pages} {218} (\bibinfo {year} {2023})}\BibitemShut {NoStop}%
\bibitem [{\citenamefont {Bian}\ \emph {et~al.}(2024{\natexlab{a}})\citenamefont {Bian}, \citenamefont {Wang}, \citenamefont {Zhang}, \citenamefont {Li}, \citenamefont {Zhang}, \citenamefont {Yang}, \citenamefont {Fang}, \citenamefont {Zhao}, \citenamefont {Zhu}, \citenamefont {Meng} \emph {et~al.}}]{bian2024broadband}%
  \BibitemOpen
  \bibfield  {author} {\bibinfo {author} {\bibfnamefont {L.}~\bibnamefont {Bian}}, \bibinfo {author} {\bibfnamefont {Z.}~\bibnamefont {Wang}}, \bibinfo {author} {\bibfnamefont {Y.}~\bibnamefont {Zhang}}, \bibinfo {author} {\bibfnamefont {L.}~\bibnamefont {Li}}, \bibinfo {author} {\bibfnamefont {Y.}~\bibnamefont {Zhang}}, \bibinfo {author} {\bibfnamefont {C.}~\bibnamefont {Yang}}, \bibinfo {author} {\bibfnamefont {W.}~\bibnamefont {Fang}}, \bibinfo {author} {\bibfnamefont {J.}~\bibnamefont {Zhao}}, \bibinfo {author} {\bibfnamefont {C.}~\bibnamefont {Zhu}}, \bibinfo {author} {\bibfnamefont {Q.}~\bibnamefont {Meng}},  \emph {et~al.},\ }\href@noop {} {\bibfield  {journal} {\bibinfo  {journal} {Nature}\ }\textbf {\bibinfo {volume} {635}},\ \bibinfo {pages} {73} (\bibinfo {year} {2024}{\natexlab{a}})}\BibitemShut {NoStop}%
\bibitem [{\citenamefont {Rubin}\ \emph {et~al.}(2019)\citenamefont {Rubin}, \citenamefont {D’Aversa}, \citenamefont {Chevalier}, \citenamefont {Shi}, \citenamefont {Chen},\ and\ \citenamefont {Capasso}}]{rubin2019matrix}%
  \BibitemOpen
  \bibfield  {author} {\bibinfo {author} {\bibfnamefont {N.~A.}\ \bibnamefont {Rubin}}, \bibinfo {author} {\bibfnamefont {G.}~\bibnamefont {D’Aversa}}, \bibinfo {author} {\bibfnamefont {P.}~\bibnamefont {Chevalier}}, \bibinfo {author} {\bibfnamefont {Z.}~\bibnamefont {Shi}}, \bibinfo {author} {\bibfnamefont {W.~T.}\ \bibnamefont {Chen}}, \ and\ \bibinfo {author} {\bibfnamefont {F.}~\bibnamefont {Capasso}},\ }\href@noop {} {\bibfield  {journal} {\bibinfo  {journal} {Science}\ }\textbf {\bibinfo {volume} {365}},\ \bibinfo {pages} {eaax1839} (\bibinfo {year} {2019})}\BibitemShut {NoStop}%
\bibitem [{\citenamefont {Gao}\ and\ \citenamefont {Wang}(2016)}]{gao2016review}%
  \BibitemOpen
  \bibfield  {author} {\bibinfo {author} {\bibfnamefont {L.}~\bibnamefont {Gao}}\ and\ \bibinfo {author} {\bibfnamefont {L.~V.}\ \bibnamefont {Wang}},\ }\href@noop {} {\bibfield  {journal} {\bibinfo  {journal} {Physics Reports}\ }\textbf {\bibinfo {volume} {616}},\ \bibinfo {pages} {1} (\bibinfo {year} {2016})}\BibitemShut {NoStop}%
\bibitem [{\citenamefont {Leonetti}\ \emph {et~al.}(2021)\citenamefont {Leonetti}, \citenamefont {H{\"o}rmann}, \citenamefont {Leuzzi}, \citenamefont {Parisi},\ and\ \citenamefont {Ruocco}}]{leonetti2021optical}%
  \BibitemOpen
  \bibfield  {author} {\bibinfo {author} {\bibfnamefont {M.}~\bibnamefont {Leonetti}}, \bibinfo {author} {\bibfnamefont {E.}~\bibnamefont {H{\"o}rmann}}, \bibinfo {author} {\bibfnamefont {L.}~\bibnamefont {Leuzzi}}, \bibinfo {author} {\bibfnamefont {G.}~\bibnamefont {Parisi}}, \ and\ \bibinfo {author} {\bibfnamefont {G.}~\bibnamefont {Ruocco}},\ }\href@noop {} {\bibfield  {journal} {\bibinfo  {journal} {Proceedings of the National Academy of Sciences}\ }\textbf {\bibinfo {volume} {118}},\ \bibinfo {pages} {e2015207118} (\bibinfo {year} {2021})}\BibitemShut {NoStop}%
\bibitem [{\citenamefont {Pierangeli}\ \emph {et~al.}(2021)\citenamefont {Pierangeli}, \citenamefont {Rafayelyan}, \citenamefont {Conti},\ and\ \citenamefont {Gigan}}]{pierangeli2021scalable}%
  \BibitemOpen
  \bibfield  {author} {\bibinfo {author} {\bibfnamefont {D.}~\bibnamefont {Pierangeli}}, \bibinfo {author} {\bibfnamefont {M.}~\bibnamefont {Rafayelyan}}, \bibinfo {author} {\bibfnamefont {C.}~\bibnamefont {Conti}}, \ and\ \bibinfo {author} {\bibfnamefont {S.}~\bibnamefont {Gigan}},\ }\href@noop {} {\bibfield  {journal} {\bibinfo  {journal} {Physical Review Applied}\ }\textbf {\bibinfo {volume} {15}},\ \bibinfo {pages} {034087} (\bibinfo {year} {2021})}\BibitemShut {NoStop}%
\bibitem [{\citenamefont {Wang}\ \emph {et~al.}(2024{\natexlab{a}})\citenamefont {Wang}, \citenamefont {Hu}, \citenamefont {Morandi}, \citenamefont {Nardi}, \citenamefont {Xia}, \citenamefont {Li}, \citenamefont {Savo}, \citenamefont {Liu}, \citenamefont {Grange},\ and\ \citenamefont {Gigan}}]{wang2024large}%
  \BibitemOpen
  \bibfield  {author} {\bibinfo {author} {\bibfnamefont {H.}~\bibnamefont {Wang}}, \bibinfo {author} {\bibfnamefont {J.}~\bibnamefont {Hu}}, \bibinfo {author} {\bibfnamefont {A.}~\bibnamefont {Morandi}}, \bibinfo {author} {\bibfnamefont {A.}~\bibnamefont {Nardi}}, \bibinfo {author} {\bibfnamefont {F.}~\bibnamefont {Xia}}, \bibinfo {author} {\bibfnamefont {X.}~\bibnamefont {Li}}, \bibinfo {author} {\bibfnamefont {R.}~\bibnamefont {Savo}}, \bibinfo {author} {\bibfnamefont {Q.}~\bibnamefont {Liu}}, \bibinfo {author} {\bibfnamefont {R.}~\bibnamefont {Grange}}, \ and\ \bibinfo {author} {\bibfnamefont {S.}~\bibnamefont {Gigan}},\ }\href@noop {} {\bibfield  {journal} {\bibinfo  {journal} {Nature Computational Science}\ }\textbf {\bibinfo {volume} {4}},\ \bibinfo {pages} {429} (\bibinfo {year} {2024}{\natexlab{a}})}\BibitemShut {NoStop}%
\bibitem [{\citenamefont {Ding}\ \emph {et~al.}(2024)\citenamefont {Ding}, \citenamefont {Shao}, \citenamefont {Li}, \citenamefont {Qu}, \citenamefont {Liu}, \citenamefont {He}, \citenamefont {Wei},\ and\ \citenamefont {Yang}}]{ding2024optoelectronic}%
  \BibitemOpen
  \bibfield  {author} {\bibinfo {author} {\bibfnamefont {C.}~\bibnamefont {Ding}}, \bibinfo {author} {\bibfnamefont {R.}~\bibnamefont {Shao}}, \bibinfo {author} {\bibfnamefont {J.}~\bibnamefont {Li}}, \bibinfo {author} {\bibfnamefont {Y.}~\bibnamefont {Qu}}, \bibinfo {author} {\bibfnamefont {L.}~\bibnamefont {Liu}}, \bibinfo {author} {\bibfnamefont {Q.}~\bibnamefont {He}}, \bibinfo {author} {\bibfnamefont {X.}~\bibnamefont {Wei}}, \ and\ \bibinfo {author} {\bibfnamefont {J.}~\bibnamefont {Yang}},\ }\href@noop {} {\bibfield  {journal} {\bibinfo  {journal} {Advanced Photonics Nexus}\ }\textbf {\bibinfo {volume} {3}},\ \bibinfo {pages} {066006} (\bibinfo {year} {2024})}\BibitemShut {NoStop}%
\bibitem [{\citenamefont {Bian}\ \emph {et~al.}(2024{\natexlab{b}})\citenamefont {Bian}, \citenamefont {Chang}, \citenamefont {Jiang}, \citenamefont {Yang}, \citenamefont {Zhan}, \citenamefont {Liu}, \citenamefont {Li}, \citenamefont {Yan}, \citenamefont {Gao},\ and\ \citenamefont {Zhang}}]{bian2024large}%
  \BibitemOpen
  \bibfield  {author} {\bibinfo {author} {\bibfnamefont {L.}~\bibnamefont {Bian}}, \bibinfo {author} {\bibfnamefont {X.}~\bibnamefont {Chang}}, \bibinfo {author} {\bibfnamefont {S.}~\bibnamefont {Jiang}}, \bibinfo {author} {\bibfnamefont {L.}~\bibnamefont {Yang}}, \bibinfo {author} {\bibfnamefont {X.}~\bibnamefont {Zhan}}, \bibinfo {author} {\bibfnamefont {S.}~\bibnamefont {Liu}}, \bibinfo {author} {\bibfnamefont {D.}~\bibnamefont {Li}}, \bibinfo {author} {\bibfnamefont {R.}~\bibnamefont {Yan}}, \bibinfo {author} {\bibfnamefont {Z.}~\bibnamefont {Gao}}, \ and\ \bibinfo {author} {\bibfnamefont {J.}~\bibnamefont {Zhang}},\ }\href@noop {} {\bibfield  {journal} {\bibinfo  {journal} {Nature Communications}\ }\textbf {\bibinfo {volume} {15}},\ \bibinfo {pages} {9807} (\bibinfo {year} {2024}{\natexlab{b}})}\BibitemShut {NoStop}%
\bibitem [{\citenamefont {Antipa}\ \emph {et~al.}(2017)\citenamefont {Antipa}, \citenamefont {Kuo}, \citenamefont {Heckel}, \citenamefont {Mildenhall}, \citenamefont {Bostan}, \citenamefont {Ng},\ and\ \citenamefont {Waller}}]{antipa2017diffusercam}%
  \BibitemOpen
  \bibfield  {author} {\bibinfo {author} {\bibfnamefont {N.}~\bibnamefont {Antipa}}, \bibinfo {author} {\bibfnamefont {G.}~\bibnamefont {Kuo}}, \bibinfo {author} {\bibfnamefont {R.}~\bibnamefont {Heckel}}, \bibinfo {author} {\bibfnamefont {B.}~\bibnamefont {Mildenhall}}, \bibinfo {author} {\bibfnamefont {E.}~\bibnamefont {Bostan}}, \bibinfo {author} {\bibfnamefont {R.}~\bibnamefont {Ng}}, \ and\ \bibinfo {author} {\bibfnamefont {L.}~\bibnamefont {Waller}},\ }\href@noop {} {\bibfield  {journal} {\bibinfo  {journal} {Optica}\ }\textbf {\bibinfo {volume} {5}},\ \bibinfo {pages} {1} (\bibinfo {year} {2017})}\BibitemShut {NoStop}%
\bibitem [{\citenamefont {Cai}\ \emph {et~al.}(2020)\citenamefont {Cai}, \citenamefont {Chen}, \citenamefont {Pedrini}, \citenamefont {Osten}, \citenamefont {Liu},\ and\ \citenamefont {Peng}}]{cai2020lensless}%
  \BibitemOpen
  \bibfield  {author} {\bibinfo {author} {\bibfnamefont {Z.}~\bibnamefont {Cai}}, \bibinfo {author} {\bibfnamefont {J.}~\bibnamefont {Chen}}, \bibinfo {author} {\bibfnamefont {G.}~\bibnamefont {Pedrini}}, \bibinfo {author} {\bibfnamefont {W.}~\bibnamefont {Osten}}, \bibinfo {author} {\bibfnamefont {X.}~\bibnamefont {Liu}}, \ and\ \bibinfo {author} {\bibfnamefont {X.}~\bibnamefont {Peng}},\ }\href@noop {} {\bibfield  {journal} {\bibinfo  {journal} {Light: Science \& Applications}\ }\textbf {\bibinfo {volume} {9}},\ \bibinfo {pages} {143} (\bibinfo {year} {2020})}\BibitemShut {NoStop}%
\bibitem [{\citenamefont {Elmalem}\ and\ \citenamefont {Giryes}(2021)}]{elmalem2021lensless}%
  \BibitemOpen
  \bibfield  {author} {\bibinfo {author} {\bibfnamefont {S.}~\bibnamefont {Elmalem}}\ and\ \bibinfo {author} {\bibfnamefont {R.}~\bibnamefont {Giryes}},\ }in\ \href@noop {} {\emph {\bibinfo {booktitle} {Computational Optical Sensing and Imaging}}}\ (\bibinfo {organization} {Optica Publishing Group},\ \bibinfo {year} {2021})\ pp.\ \bibinfo {pages} {CTh7A--1}\BibitemShut {NoStop}%
\bibitem [{\citenamefont {Baek}\ \emph {et~al.}(2022)\citenamefont {Baek}, \citenamefont {Lee}, \citenamefont {Kim}, \citenamefont {Jung},\ and\ \citenamefont {Lee}}]{baek2022lensless}%
  \BibitemOpen
  \bibfield  {author} {\bibinfo {author} {\bibfnamefont {N.}~\bibnamefont {Baek}}, \bibinfo {author} {\bibfnamefont {Y.}~\bibnamefont {Lee}}, \bibinfo {author} {\bibfnamefont {T.}~\bibnamefont {Kim}}, \bibinfo {author} {\bibfnamefont {J.}~\bibnamefont {Jung}}, \ and\ \bibinfo {author} {\bibfnamefont {S.~A.}\ \bibnamefont {Lee}},\ }\href@noop {} {\bibfield  {journal} {\bibinfo  {journal} {APL Photonics}\ }\textbf {\bibinfo {volume} {7}} (\bibinfo {year} {2022})}\BibitemShut {NoStop}%
\bibitem [{\citenamefont {Pierangeli}\ \emph {et~al.}(2024)\citenamefont {Pierangeli}, \citenamefont {Volpe},\ and\ \citenamefont {Conti}}]{pierangeli2024deep}%
  \BibitemOpen
  \bibfield  {author} {\bibinfo {author} {\bibfnamefont {D.}~\bibnamefont {Pierangeli}}, \bibinfo {author} {\bibfnamefont {G.}~\bibnamefont {Volpe}}, \ and\ \bibinfo {author} {\bibfnamefont {C.}~\bibnamefont {Conti}},\ }\href@noop {} {\bibfield  {journal} {\bibinfo  {journal} {Laser \& Photonics Reviews}\ }\textbf {\bibinfo {volume} {18}},\ \bibinfo {pages} {2400626} (\bibinfo {year} {2024})}\BibitemShut {NoStop}%
\bibitem [{\citenamefont {Sahoo}\ \emph {et~al.}(2017)\citenamefont {Sahoo}, \citenamefont {Tang},\ and\ \citenamefont {Dang}}]{sahoo2017single}%
  \BibitemOpen
  \bibfield  {author} {\bibinfo {author} {\bibfnamefont {S.~K.}\ \bibnamefont {Sahoo}}, \bibinfo {author} {\bibfnamefont {D.}~\bibnamefont {Tang}}, \ and\ \bibinfo {author} {\bibfnamefont {C.}~\bibnamefont {Dang}},\ }\href@noop {} {\bibfield  {journal} {\bibinfo  {journal} {Optica}\ }\textbf {\bibinfo {volume} {4}},\ \bibinfo {pages} {1209} (\bibinfo {year} {2017})}\BibitemShut {NoStop}%
\bibitem [{\citenamefont {Li}\ \emph {et~al.}(2019)\citenamefont {Li}, \citenamefont {Greenberg},\ and\ \citenamefont {Gehm}}]{li2019single}%
  \BibitemOpen
  \bibfield  {author} {\bibinfo {author} {\bibfnamefont {X.}~\bibnamefont {Li}}, \bibinfo {author} {\bibfnamefont {J.~A.}\ \bibnamefont {Greenberg}}, \ and\ \bibinfo {author} {\bibfnamefont {M.~E.}\ \bibnamefont {Gehm}},\ }\href@noop {} {\bibfield  {journal} {\bibinfo  {journal} {Optica}\ }\textbf {\bibinfo {volume} {6}},\ \bibinfo {pages} {864} (\bibinfo {year} {2019})}\BibitemShut {NoStop}%
\bibitem [{\citenamefont {Li}\ \emph {et~al.}(2022)\citenamefont {Li}, \citenamefont {Wei}, \citenamefont {Li}, \citenamefont {Zhu}, \citenamefont {Peng},\ and\ \citenamefont {Ma}}]{li2022hyperspectral}%
  \BibitemOpen
  \bibfield  {author} {\bibinfo {author} {\bibfnamefont {Y.}~\bibnamefont {Li}}, \bibinfo {author} {\bibfnamefont {S.}~\bibnamefont {Wei}}, \bibinfo {author} {\bibfnamefont {Z.}~\bibnamefont {Li}}, \bibinfo {author} {\bibfnamefont {Z.}~\bibnamefont {Zhu}}, \bibinfo {author} {\bibfnamefont {J.}~\bibnamefont {Peng}}, \ and\ \bibinfo {author} {\bibfnamefont {D.}~\bibnamefont {Ma}},\ }\href@noop {} {\bibfield  {journal} {\bibinfo  {journal} {Applied Physics Letters}\ }\textbf {\bibinfo {volume} {120}} (\bibinfo {year} {2022})}\BibitemShut {NoStop}%
\bibitem [{\citenamefont {Kim}\ \emph {et~al.}(2023)\citenamefont {Kim}, \citenamefont {Lee}, \citenamefont {Baek}, \citenamefont {Chae},\ and\ \citenamefont {Lee}}]{kim2023aperture}%
  \BibitemOpen
  \bibfield  {author} {\bibinfo {author} {\bibfnamefont {T.}~\bibnamefont {Kim}}, \bibinfo {author} {\bibfnamefont {K.~C.}\ \bibnamefont {Lee}}, \bibinfo {author} {\bibfnamefont {N.}~\bibnamefont {Baek}}, \bibinfo {author} {\bibfnamefont {H.}~\bibnamefont {Chae}}, \ and\ \bibinfo {author} {\bibfnamefont {S.~A.}\ \bibnamefont {Lee}},\ }\href@noop {} {\bibfield  {journal} {\bibinfo  {journal} {APL Photonics}\ }\textbf {\bibinfo {volume} {8}} (\bibinfo {year} {2023})}\BibitemShut {NoStop}%
\bibitem [{\citenamefont {Malone}\ \emph {et~al.}(2023)\citenamefont {Malone}, \citenamefont {Aggarwal}, \citenamefont {Waller},\ and\ \citenamefont {Bowden}}]{malone2023diffuserspec}%
  \BibitemOpen
  \bibfield  {author} {\bibinfo {author} {\bibfnamefont {J.~D.}\ \bibnamefont {Malone}}, \bibinfo {author} {\bibfnamefont {N.}~\bibnamefont {Aggarwal}}, \bibinfo {author} {\bibfnamefont {L.}~\bibnamefont {Waller}}, \ and\ \bibinfo {author} {\bibfnamefont {A.~K.}\ \bibnamefont {Bowden}},\ }\href@noop {} {\bibfield  {journal} {\bibinfo  {journal} {Optics Letters}\ }\textbf {\bibinfo {volume} {48}},\ \bibinfo {pages} {323} (\bibinfo {year} {2023})}\BibitemShut {NoStop}%
\bibitem [{\citenamefont {Monakhova}\ \emph {et~al.}(2020)\citenamefont {Monakhova}, \citenamefont {Yanny}, \citenamefont {Aggarwal},\ and\ \citenamefont {Waller}}]{monakhova2020spectral}%
  \BibitemOpen
  \bibfield  {author} {\bibinfo {author} {\bibfnamefont {K.}~\bibnamefont {Monakhova}}, \bibinfo {author} {\bibfnamefont {K.}~\bibnamefont {Yanny}}, \bibinfo {author} {\bibfnamefont {N.}~\bibnamefont {Aggarwal}}, \ and\ \bibinfo {author} {\bibfnamefont {L.}~\bibnamefont {Waller}},\ }\href@noop {} {\bibfield  {journal} {\bibinfo  {journal} {Optica}\ }\textbf {\bibinfo {volume} {7}},\ \bibinfo {pages} {1298} (\bibinfo {year} {2020})}\BibitemShut {NoStop}%
\bibitem [{\citenamefont {Liu}\ \emph {et~al.}(2021{\natexlab{a}})\citenamefont {Liu}, \citenamefont {Wang}, \citenamefont {Chen},\ and\ \citenamefont {McGloin}}]{liu2021single}%
  \BibitemOpen
  \bibfield  {author} {\bibinfo {author} {\bibfnamefont {B.}~\bibnamefont {Liu}}, \bibinfo {author} {\bibfnamefont {F.}~\bibnamefont {Wang}}, \bibinfo {author} {\bibfnamefont {C.}~\bibnamefont {Chen}}, \ and\ \bibinfo {author} {\bibfnamefont {D.}~\bibnamefont {McGloin}},\ }\href@noop {} {\bibfield  {journal} {\bibinfo  {journal} {IEEE Photonics Journal}\ }\textbf {\bibinfo {volume} {13}},\ \bibinfo {pages} {1} (\bibinfo {year} {2021}{\natexlab{a}})}\BibitemShut {NoStop}%
\bibitem [{\citenamefont {Adams}\ \emph {et~al.}(2022)\citenamefont {Adams}, \citenamefont {Yan}, \citenamefont {Wu}, \citenamefont {Boominathan}, \citenamefont {Gao}, \citenamefont {Rodriguez}, \citenamefont {Kim}, \citenamefont {Carns}, \citenamefont {Richards-Kortum}, \citenamefont {Kemere} \emph {et~al.}}]{adams2022vivo}%
  \BibitemOpen
  \bibfield  {author} {\bibinfo {author} {\bibfnamefont {J.~K.}\ \bibnamefont {Adams}}, \bibinfo {author} {\bibfnamefont {D.}~\bibnamefont {Yan}}, \bibinfo {author} {\bibfnamefont {J.}~\bibnamefont {Wu}}, \bibinfo {author} {\bibfnamefont {V.}~\bibnamefont {Boominathan}}, \bibinfo {author} {\bibfnamefont {S.}~\bibnamefont {Gao}}, \bibinfo {author} {\bibfnamefont {A.~V.}\ \bibnamefont {Rodriguez}}, \bibinfo {author} {\bibfnamefont {S.}~\bibnamefont {Kim}}, \bibinfo {author} {\bibfnamefont {J.}~\bibnamefont {Carns}}, \bibinfo {author} {\bibfnamefont {R.}~\bibnamefont {Richards-Kortum}}, \bibinfo {author} {\bibfnamefont {C.}~\bibnamefont {Kemere}},  \emph {et~al.},\ }\href@noop {} {\bibfield  {journal} {\bibinfo  {journal} {Nature Biomedical Engineering}\ }\textbf {\bibinfo {volume} {6}},\ \bibinfo {pages} {617} (\bibinfo {year} {2022})}\BibitemShut {NoStop}%
\bibitem [{\citenamefont {Wu}\ \emph {et~al.}(2024{\natexlab{a}})\citenamefont {Wu}, \citenamefont {Chen}, \citenamefont {Veeraraghavan}, \citenamefont {Seidemann},\ and\ \citenamefont {Robinson}}]{wu2024mesoscopic}%
  \BibitemOpen
  \bibfield  {author} {\bibinfo {author} {\bibfnamefont {J.}~\bibnamefont {Wu}}, \bibinfo {author} {\bibfnamefont {Y.}~\bibnamefont {Chen}}, \bibinfo {author} {\bibfnamefont {A.}~\bibnamefont {Veeraraghavan}}, \bibinfo {author} {\bibfnamefont {E.}~\bibnamefont {Seidemann}}, \ and\ \bibinfo {author} {\bibfnamefont {J.~T.}\ \bibnamefont {Robinson}},\ }\href@noop {} {\bibfield  {journal} {\bibinfo  {journal} {Nature Communications}\ }\textbf {\bibinfo {volume} {15}},\ \bibinfo {pages} {1271} (\bibinfo {year} {2024}{\natexlab{a}})}\BibitemShut {NoStop}%
\bibitem [{\citenamefont {Zheng}\ \emph {et~al.}(2024)\citenamefont {Zheng}, \citenamefont {Liu}, \citenamefont {Song}, \citenamefont {Ding}, \citenamefont {Zhong}, \citenamefont {Chang}, \citenamefont {Wu}, \citenamefont {McGloin},\ and\ \citenamefont {Wang}}]{zheng2024temporal}%
  \BibitemOpen
  \bibfield  {author} {\bibinfo {author} {\bibfnamefont {Z.}~\bibnamefont {Zheng}}, \bibinfo {author} {\bibfnamefont {B.}~\bibnamefont {Liu}}, \bibinfo {author} {\bibfnamefont {J.}~\bibnamefont {Song}}, \bibinfo {author} {\bibfnamefont {L.}~\bibnamefont {Ding}}, \bibinfo {author} {\bibfnamefont {X.}~\bibnamefont {Zhong}}, \bibinfo {author} {\bibfnamefont {L.}~\bibnamefont {Chang}}, \bibinfo {author} {\bibfnamefont {X.}~\bibnamefont {Wu}}, \bibinfo {author} {\bibfnamefont {D.}~\bibnamefont {McGloin}}, \ and\ \bibinfo {author} {\bibfnamefont {F.}~\bibnamefont {Wang}},\ }\href@noop {} {\bibfield  {journal} {\bibinfo  {journal} {Optics Letters}\ }\textbf {\bibinfo {volume} {49}},\ \bibinfo {pages} {3058} (\bibinfo {year} {2024})}\BibitemShut {NoStop}%
\bibitem [{\citenamefont {Antipa}\ \emph {et~al.}(2019)\citenamefont {Antipa}, \citenamefont {Oare}, \citenamefont {Bostan}, \citenamefont {Ng},\ and\ \citenamefont {Waller}}]{antipa2019video}%
  \BibitemOpen
  \bibfield  {author} {\bibinfo {author} {\bibfnamefont {N.}~\bibnamefont {Antipa}}, \bibinfo {author} {\bibfnamefont {P.}~\bibnamefont {Oare}}, \bibinfo {author} {\bibfnamefont {E.}~\bibnamefont {Bostan}}, \bibinfo {author} {\bibfnamefont {R.}~\bibnamefont {Ng}}, \ and\ \bibinfo {author} {\bibfnamefont {L.}~\bibnamefont {Waller}},\ }in\ \href@noop {} {\emph {\bibinfo {booktitle} {2019 IEEE International Conference on Computational Photography (ICCP)}}}\ (\bibinfo {organization} {IEEE},\ \bibinfo {year} {2019})\ pp.\ \bibinfo {pages} {1--8}\BibitemShut {NoStop}%
\bibitem [{\citenamefont {Weinberg}\ and\ \citenamefont {Katz}(2020)}]{weinberg2020100}%
  \BibitemOpen
  \bibfield  {author} {\bibinfo {author} {\bibfnamefont {G.}~\bibnamefont {Weinberg}}\ and\ \bibinfo {author} {\bibfnamefont {O.}~\bibnamefont {Katz}},\ }\href@noop {} {\bibfield  {journal} {\bibinfo  {journal} {Optics Express}\ }\textbf {\bibinfo {volume} {28}},\ \bibinfo {pages} {30616} (\bibinfo {year} {2020})}\BibitemShut {NoStop}%
\bibitem [{\citenamefont {Wu}\ \emph {et~al.}(2024{\natexlab{b}})\citenamefont {Wu}, \citenamefont {Guillon}, \citenamefont {Tessier},\ and\ \citenamefont {Berto}}]{wu2024multiplexed}%
  \BibitemOpen
  \bibfield  {author} {\bibinfo {author} {\bibfnamefont {T.}~\bibnamefont {Wu}}, \bibinfo {author} {\bibfnamefont {M.}~\bibnamefont {Guillon}}, \bibinfo {author} {\bibfnamefont {G.}~\bibnamefont {Tessier}}, \ and\ \bibinfo {author} {\bibfnamefont {P.}~\bibnamefont {Berto}},\ }\href@noop {} {\bibfield  {journal} {\bibinfo  {journal} {Optica}\ }\textbf {\bibinfo {volume} {11}},\ \bibinfo {pages} {297} (\bibinfo {year} {2024}{\natexlab{b}})}\BibitemShut {NoStop}%
\bibitem [{\citenamefont {Zhu}\ \emph {et~al.}(2023)\citenamefont {Zhu}, \citenamefont {Zheng}, \citenamefont {Meng}, \citenamefont {Chang}, \citenamefont {Tan}, \citenamefont {Chen}, \citenamefont {Fang}, \citenamefont {Gu},\ and\ \citenamefont {Chen}}]{zhu2023harnessing}%
  \BibitemOpen
  \bibfield  {author} {\bibinfo {author} {\bibfnamefont {S.-k.}\ \bibnamefont {Zhu}}, \bibinfo {author} {\bibfnamefont {Z.-h.}\ \bibnamefont {Zheng}}, \bibinfo {author} {\bibfnamefont {W.}~\bibnamefont {Meng}}, \bibinfo {author} {\bibfnamefont {S.-s.}\ \bibnamefont {Chang}}, \bibinfo {author} {\bibfnamefont {Y.}~\bibnamefont {Tan}}, \bibinfo {author} {\bibfnamefont {L.-J.}\ \bibnamefont {Chen}}, \bibinfo {author} {\bibfnamefont {X.}~\bibnamefont {Fang}}, \bibinfo {author} {\bibfnamefont {M.}~\bibnamefont {Gu}}, \ and\ \bibinfo {author} {\bibfnamefont {J.-h.}\ \bibnamefont {Chen}},\ }\href@noop {} {\bibfield  {journal} {\bibinfo  {journal} {PhotoniX}\ }\textbf {\bibinfo {volume} {4}},\ \bibinfo {pages} {26} (\bibinfo {year} {2023})}\BibitemShut {NoStop}%
\bibitem [{\citenamefont {Sun}\ \emph {et~al.}(2024)\citenamefont {Sun}, \citenamefont {Nie}, \citenamefont {Du}, \citenamefont {Chang},\ and\ \citenamefont {Liu}}]{sun2024overcoming}%
  \BibitemOpen
  \bibfield  {author} {\bibinfo {author} {\bibfnamefont {S.}~\bibnamefont {Sun}}, \bibinfo {author} {\bibfnamefont {Z.-W.}\ \bibnamefont {Nie}}, \bibinfo {author} {\bibfnamefont {L.-K.}\ \bibnamefont {Du}}, \bibinfo {author} {\bibfnamefont {C.}~\bibnamefont {Chang}}, \ and\ \bibinfo {author} {\bibfnamefont {W.-T.}\ \bibnamefont {Liu}},\ }\href@noop {} {\bibfield  {journal} {\bibinfo  {journal} {Optica}\ }\textbf {\bibinfo {volume} {11}},\ \bibinfo {pages} {385} (\bibinfo {year} {2024})}\BibitemShut {NoStop}%
\bibitem [{\citenamefont {Tian}\ \emph {et~al.}(2024{\natexlab{a}})\citenamefont {Tian}, \citenamefont {Li}, \citenamefont {Ma}, \citenamefont {Cao},\ and\ \citenamefont {Su}}]{tian2024cfza}%
  \BibitemOpen
  \bibfield  {author} {\bibinfo {author} {\bibfnamefont {Z.}~\bibnamefont {Tian}}, \bibinfo {author} {\bibfnamefont {L.}~\bibnamefont {Li}}, \bibinfo {author} {\bibfnamefont {J.}~\bibnamefont {Ma}}, \bibinfo {author} {\bibfnamefont {L.}~\bibnamefont {Cao}}, \ and\ \bibinfo {author} {\bibfnamefont {P.}~\bibnamefont {Su}},\ }\href@noop {} {\bibfield  {journal} {\bibinfo  {journal} {Optics Letters}\ }\textbf {\bibinfo {volume} {49}},\ \bibinfo {pages} {3532} (\bibinfo {year} {2024}{\natexlab{a}})}\BibitemShut {NoStop}%
\bibitem [{\citenamefont {Shen}\ \emph {et~al.}(2023)\citenamefont {Shen}, \citenamefont {Zhao}, \citenamefont {Jin}, \citenamefont {Wang}, \citenamefont {Cao},\ and\ \citenamefont {Yang}}]{shen2023monocular}%
  \BibitemOpen
  \bibfield  {author} {\bibinfo {author} {\bibfnamefont {Z.}~\bibnamefont {Shen}}, \bibinfo {author} {\bibfnamefont {F.}~\bibnamefont {Zhao}}, \bibinfo {author} {\bibfnamefont {C.}~\bibnamefont {Jin}}, \bibinfo {author} {\bibfnamefont {S.}~\bibnamefont {Wang}}, \bibinfo {author} {\bibfnamefont {L.}~\bibnamefont {Cao}}, \ and\ \bibinfo {author} {\bibfnamefont {Y.}~\bibnamefont {Yang}},\ }\href@noop {} {\bibfield  {journal} {\bibinfo  {journal} {Nature Communications}\ }\textbf {\bibinfo {volume} {14}},\ \bibinfo {pages} {1035} (\bibinfo {year} {2023})}\BibitemShut {NoStop}%
\bibitem [{\citenamefont {Lei}\ \emph {et~al.}(2023)\citenamefont {Lei}, \citenamefont {Zhang}, \citenamefont {Guo}, \citenamefont {Pu}, \citenamefont {Zou}, \citenamefont {Li}, \citenamefont {Ma},\ and\ \citenamefont {Luo}}]{lei2023snapshot}%
  \BibitemOpen
  \bibfield  {author} {\bibinfo {author} {\bibfnamefont {Y.}~\bibnamefont {Lei}}, \bibinfo {author} {\bibfnamefont {Q.}~\bibnamefont {Zhang}}, \bibinfo {author} {\bibfnamefont {Y.}~\bibnamefont {Guo}}, \bibinfo {author} {\bibfnamefont {M.}~\bibnamefont {Pu}}, \bibinfo {author} {\bibfnamefont {F.}~\bibnamefont {Zou}}, \bibinfo {author} {\bibfnamefont {X.}~\bibnamefont {Li}}, \bibinfo {author} {\bibfnamefont {X.}~\bibnamefont {Ma}}, \ and\ \bibinfo {author} {\bibfnamefont {X.}~\bibnamefont {Luo}},\ }\href@noop {} {\bibfield  {journal} {\bibinfo  {journal} {Photonics Research}\ }\textbf {\bibinfo {volume} {11}},\ \bibinfo {pages} {B111} (\bibinfo {year} {2023})}\BibitemShut {NoStop}%
\bibitem [{\citenamefont {Miller}\ \emph {et~al.}(2020)\citenamefont {Miller}, \citenamefont {Wang}, \citenamefont {Keating},\ and\ \citenamefont {Liu}}]{miller2020particle}%
  \BibitemOpen
  \bibfield  {author} {\bibinfo {author} {\bibfnamefont {J.~R.}\ \bibnamefont {Miller}}, \bibinfo {author} {\bibfnamefont {C.-Y.}\ \bibnamefont {Wang}}, \bibinfo {author} {\bibfnamefont {C.~D.}\ \bibnamefont {Keating}}, \ and\ \bibinfo {author} {\bibfnamefont {Z.}~\bibnamefont {Liu}},\ }\href@noop {} {\bibfield  {journal} {\bibinfo  {journal} {ACS Nano}\ }\textbf {\bibinfo {volume} {14}},\ \bibinfo {pages} {13038} (\bibinfo {year} {2020})}\BibitemShut {NoStop}%
\bibitem [{\citenamefont {Zhang}\ \emph {et~al.}(2025)\citenamefont {Zhang}, \citenamefont {Wang}, \citenamefont {Li}, \citenamefont {Liang}, \citenamefont {Guan}, \citenamefont {Chen},\ and\ \citenamefont {Zuo}}]{zhang2025lensless}%
  \BibitemOpen
  \bibfield  {author} {\bibinfo {author} {\bibfnamefont {X.}~\bibnamefont {Zhang}}, \bibinfo {author} {\bibfnamefont {B.}~\bibnamefont {Wang}}, \bibinfo {author} {\bibfnamefont {S.}~\bibnamefont {Li}}, \bibinfo {author} {\bibfnamefont {K.}~\bibnamefont {Liang}}, \bibinfo {author} {\bibfnamefont {H.}~\bibnamefont {Guan}}, \bibinfo {author} {\bibfnamefont {Q.}~\bibnamefont {Chen}}, \ and\ \bibinfo {author} {\bibfnamefont {C.}~\bibnamefont {Zuo}},\ }\href@noop {} {\bibfield  {journal} {\bibinfo  {journal} {Science Advances}\ }\textbf {\bibinfo {volume} {11}},\ \bibinfo {pages} {eadt3909} (\bibinfo {year} {2025})}\BibitemShut {NoStop}%
\bibitem [{\citenamefont {Donoho}(2006)}]{donoho2006compressed}%
  \BibitemOpen
  \bibfield  {author} {\bibinfo {author} {\bibfnamefont {D.~L.}\ \bibnamefont {Donoho}},\ }\href@noop {} {\bibfield  {journal} {\bibinfo  {journal} {IEEE Transactions on Information Theory}\ }\textbf {\bibinfo {volume} {52}},\ \bibinfo {pages} {1289} (\bibinfo {year} {2006})}\BibitemShut {NoStop}%
\bibitem [{\citenamefont {Liu}\ \emph {et~al.}(2021{\natexlab{b}})\citenamefont {Liu}, \citenamefont {Wang}, \citenamefont {Chen}, \citenamefont {Dong},\ and\ \citenamefont {McGloin}}]{liu2021self}%
  \BibitemOpen
  \bibfield  {author} {\bibinfo {author} {\bibfnamefont {B.}~\bibnamefont {Liu}}, \bibinfo {author} {\bibfnamefont {F.}~\bibnamefont {Wang}}, \bibinfo {author} {\bibfnamefont {C.}~\bibnamefont {Chen}}, \bibinfo {author} {\bibfnamefont {F.}~\bibnamefont {Dong}}, \ and\ \bibinfo {author} {\bibfnamefont {D.}~\bibnamefont {McGloin}},\ }\href@noop {} {\bibfield  {journal} {\bibinfo  {journal} {Optica}\ }\textbf {\bibinfo {volume} {8}},\ \bibinfo {pages} {1340} (\bibinfo {year} {2021}{\natexlab{b}})}\BibitemShut {NoStop}%
\bibitem [{\citenamefont {Wang}\ \emph {et~al.}(2023)\citenamefont {Wang}, \citenamefont {Liu}, \citenamefont {Song}, \citenamefont {Wang}, \citenamefont {Shan}, \citenamefont {Zhong},\ and\ \citenamefont {Wang}}]{wang2023dual}%
  \BibitemOpen
  \bibfield  {author} {\bibinfo {author} {\bibfnamefont {D.}~\bibnamefont {Wang}}, \bibinfo {author} {\bibfnamefont {B.}~\bibnamefont {Liu}}, \bibinfo {author} {\bibfnamefont {J.}~\bibnamefont {Song}}, \bibinfo {author} {\bibfnamefont {Y.}~\bibnamefont {Wang}}, \bibinfo {author} {\bibfnamefont {X.}~\bibnamefont {Shan}}, \bibinfo {author} {\bibfnamefont {X.}~\bibnamefont {Zhong}}, \ and\ \bibinfo {author} {\bibfnamefont {F.}~\bibnamefont {Wang}},\ }\href@noop {} {\bibfield  {journal} {\bibinfo  {journal} {Optics Express}\ }\textbf {\bibinfo {volume} {31}},\ \bibinfo {pages} {14225} (\bibinfo {year} {2023})}\BibitemShut {NoStop}%
\bibitem [{\citenamefont {Song}\ \emph {et~al.}(2024)\citenamefont {Song}, \citenamefont {Liu}, \citenamefont {Wang}, \citenamefont {Chen}, \citenamefont {Shan}, \citenamefont {Zhong}, \citenamefont {Wu},\ and\ \citenamefont {Wang}}]{song2024computational}%
  \BibitemOpen
  \bibfield  {author} {\bibinfo {author} {\bibfnamefont {J.}~\bibnamefont {Song}}, \bibinfo {author} {\bibfnamefont {B.}~\bibnamefont {Liu}}, \bibinfo {author} {\bibfnamefont {Y.}~\bibnamefont {Wang}}, \bibinfo {author} {\bibfnamefont {C.}~\bibnamefont {Chen}}, \bibinfo {author} {\bibfnamefont {X.}~\bibnamefont {Shan}}, \bibinfo {author} {\bibfnamefont {X.}~\bibnamefont {Zhong}}, \bibinfo {author} {\bibfnamefont {L.-A.}\ \bibnamefont {Wu}}, \ and\ \bibinfo {author} {\bibfnamefont {F.}~\bibnamefont {Wang}},\ }\href@noop {} {\bibfield  {journal} {\bibinfo  {journal} {Photonics Research}\ }\textbf {\bibinfo {volume} {12}},\ \bibinfo {pages} {226} (\bibinfo {year} {2024})}\BibitemShut {NoStop}%
\bibitem [{\citenamefont {Bioucas-Dias}\ and\ \citenamefont {Figueiredo}(2007)}]{bioucas2007new}%
  \BibitemOpen
  \bibfield  {author} {\bibinfo {author} {\bibfnamefont {J.~M.}\ \bibnamefont {Bioucas-Dias}}\ and\ \bibinfo {author} {\bibfnamefont {M.~A.}\ \bibnamefont {Figueiredo}},\ }\href@noop {} {\bibfield  {journal} {\bibinfo  {journal} {IEEE Transactions on Image Processing}\ }\textbf {\bibinfo {volume} {16}},\ \bibinfo {pages} {2992} (\bibinfo {year} {2007})}\BibitemShut {NoStop}%
\bibitem [{\citenamefont {Kabuli}\ \emph {et~al.}(2025)\citenamefont {Kabuli}, \citenamefont {Pinkard}, \citenamefont {Markley}, \citenamefont {Hung},\ and\ \citenamefont {Waller}}]{kabuli2025designinglenslessimagingsystems}%
  \BibitemOpen
  \bibfield  {author} {\bibinfo {author} {\bibfnamefont {L.~A.}\ \bibnamefont {Kabuli}}, \bibinfo {author} {\bibfnamefont {H.}~\bibnamefont {Pinkard}}, \bibinfo {author} {\bibfnamefont {E.}~\bibnamefont {Markley}}, \bibinfo {author} {\bibfnamefont {C.~S.}\ \bibnamefont {Hung}}, \ and\ \bibinfo {author} {\bibfnamefont {L.}~\bibnamefont {Waller}},\ }\href@noop {} {\bibfield  {journal} {\bibinfo  {journal} {arXiv}\ ,\ \bibinfo {pages} {2506.08513}} (\bibinfo {year} {2025})}\BibitemShut {NoStop}%
\bibitem [{\citenamefont {Boominathan}\ \emph {et~al.}(2021)\citenamefont {Boominathan}, \citenamefont {Robinson}, \citenamefont {Waller},\ and\ \citenamefont {Veeraraghavan}}]{boominathan2021recent}%
  \BibitemOpen
  \bibfield  {author} {\bibinfo {author} {\bibfnamefont {V.}~\bibnamefont {Boominathan}}, \bibinfo {author} {\bibfnamefont {J.~T.}\ \bibnamefont {Robinson}}, \bibinfo {author} {\bibfnamefont {L.}~\bibnamefont {Waller}}, \ and\ \bibinfo {author} {\bibfnamefont {A.}~\bibnamefont {Veeraraghavan}},\ }\href@noop {} {\bibfield  {journal} {\bibinfo  {journal} {Optica}\ }\textbf {\bibinfo {volume} {9}},\ \bibinfo {pages} {1} (\bibinfo {year} {2021})}\BibitemShut {NoStop}%
\bibitem [{\citenamefont {Zhang}\ \emph {et~al.}(2024)\citenamefont {Zhang}, \citenamefont {Qiu}, \citenamefont {Zhao}, \citenamefont {Liu}, \citenamefont {Zhao},\ and\ \citenamefont {Gao}}]{zhang2024robust}%
  \BibitemOpen
  \bibfield  {author} {\bibinfo {author} {\bibfnamefont {A.}~\bibnamefont {Zhang}}, \bibinfo {author} {\bibfnamefont {S.}~\bibnamefont {Qiu}}, \bibinfo {author} {\bibfnamefont {L.}~\bibnamefont {Zhao}}, \bibinfo {author} {\bibfnamefont {H.}~\bibnamefont {Liu}}, \bibinfo {author} {\bibfnamefont {J.}~\bibnamefont {Zhao}}, \ and\ \bibinfo {author} {\bibfnamefont {J.}~\bibnamefont {Gao}},\ }\href@noop {} {\bibfield  {journal} {\bibinfo  {journal} {Laser \& Photonics Reviews}\ }\textbf {\bibinfo {volume} {18}},\ \bibinfo {pages} {2300742} (\bibinfo {year} {2024})}\BibitemShut {NoStop}%
\bibitem [{\citenamefont {Wang}\ \emph {et~al.}(2024{\natexlab{b}})\citenamefont {Wang}, \citenamefont {Liu}, \citenamefont {Ding}, \citenamefont {Chen}, \citenamefont {Shan}, \citenamefont {Wang}, \citenamefont {Tian}, \citenamefont {Song}, \citenamefont {Zheng}, \citenamefont {Xu} \emph {et~al.}}]{wang2024multi}%
  \BibitemOpen
  \bibfield  {author} {\bibinfo {author} {\bibfnamefont {Y.}~\bibnamefont {Wang}}, \bibinfo {author} {\bibfnamefont {B.}~\bibnamefont {Liu}}, \bibinfo {author} {\bibfnamefont {L.}~\bibnamefont {Ding}}, \bibinfo {author} {\bibfnamefont {C.}~\bibnamefont {Chen}}, \bibinfo {author} {\bibfnamefont {X.}~\bibnamefont {Shan}}, \bibinfo {author} {\bibfnamefont {D.}~\bibnamefont {Wang}}, \bibinfo {author} {\bibfnamefont {M.}~\bibnamefont {Tian}}, \bibinfo {author} {\bibfnamefont {J.}~\bibnamefont {Song}}, \bibinfo {author} {\bibfnamefont {Z.}~\bibnamefont {Zheng}}, \bibinfo {author} {\bibfnamefont {X.}~\bibnamefont {Xu}},  \emph {et~al.},\ }\href@noop {} {\bibfield  {journal} {\bibinfo  {journal} {Laser \& Photonics Reviews}\ }\textbf {\bibinfo {volume} {18}},\ \bibinfo {pages} {2400746} (\bibinfo {year} {2024}{\natexlab{b}})}\BibitemShut {NoStop}%
\bibitem [{\citenamefont {Tian}\ \emph {et~al.}(2024{\natexlab{b}})\citenamefont {Tian}, \citenamefont {Liu}, \citenamefont {Lu}, \citenamefont {Wang}, \citenamefont {Zheng}, \citenamefont {Song}, \citenamefont {Zhong},\ and\ \citenamefont {Wang}}]{tian2024miniaturized}%
  \BibitemOpen
  \bibfield  {author} {\bibinfo {author} {\bibfnamefont {M.}~\bibnamefont {Tian}}, \bibinfo {author} {\bibfnamefont {B.}~\bibnamefont {Liu}}, \bibinfo {author} {\bibfnamefont {Z.}~\bibnamefont {Lu}}, \bibinfo {author} {\bibfnamefont {Y.}~\bibnamefont {Wang}}, \bibinfo {author} {\bibfnamefont {Z.}~\bibnamefont {Zheng}}, \bibinfo {author} {\bibfnamefont {J.}~\bibnamefont {Song}}, \bibinfo {author} {\bibfnamefont {X.}~\bibnamefont {Zhong}}, \ and\ \bibinfo {author} {\bibfnamefont {F.}~\bibnamefont {Wang}},\ }\href@noop {} {\bibfield  {journal} {\bibinfo  {journal} {Light: Science \& Applications}\ }\textbf {\bibinfo {volume} {13}},\ \bibinfo {pages} {278} (\bibinfo {year} {2024}{\natexlab{b}})}\BibitemShut {NoStop}%
\bibitem [{\citenamefont {Gao}\ and\ \citenamefont {Cao}(2024)}]{gao2024motion}%
  \BibitemOpen
  \bibfield  {author} {\bibinfo {author} {\bibfnamefont {Y.}~\bibnamefont {Gao}}\ and\ \bibinfo {author} {\bibfnamefont {L.}~\bibnamefont {Cao}},\ }\href@noop {} {\bibfield  {journal} {\bibinfo  {journal} {Optica}\ }\textbf {\bibinfo {volume} {11}},\ \bibinfo {pages} {32} (\bibinfo {year} {2024})}\BibitemShut {NoStop}%
\bibitem [{\citenamefont {Cao}\ \emph {et~al.}(2024)\citenamefont {Cao}, \citenamefont {Divekar}, \citenamefont {Nu{\~n}ez}, \citenamefont {Upadhyayula},\ and\ \citenamefont {Waller}}]{cao2024neural}%
  \BibitemOpen
  \bibfield  {author} {\bibinfo {author} {\bibfnamefont {R.}~\bibnamefont {Cao}}, \bibinfo {author} {\bibfnamefont {N.~S.}\ \bibnamefont {Divekar}}, \bibinfo {author} {\bibfnamefont {J.~K.}\ \bibnamefont {Nu{\~n}ez}}, \bibinfo {author} {\bibfnamefont {S.}~\bibnamefont {Upadhyayula}}, \ and\ \bibinfo {author} {\bibfnamefont {L.}~\bibnamefont {Waller}},\ }\href@noop {} {\bibfield  {journal} {\bibinfo  {journal} {Nature Methods}\ }\textbf {\bibinfo {volume} {21}},\ \bibinfo {pages} {2336} (\bibinfo {year} {2024})}\BibitemShut {NoStop}%
\bibitem [{\citenamefont {Xiao}\ \emph {et~al.}(2024)\citenamefont {Xiao}, \citenamefont {Zhai}, \citenamefont {Huang}, \citenamefont {Fan},\ and\ \citenamefont {Zeng}}]{xiao2024quantum}%
  \BibitemOpen
  \bibfield  {author} {\bibinfo {author} {\bibfnamefont {T.}~\bibnamefont {Xiao}}, \bibinfo {author} {\bibfnamefont {X.}~\bibnamefont {Zhai}}, \bibinfo {author} {\bibfnamefont {J.}~\bibnamefont {Huang}}, \bibinfo {author} {\bibfnamefont {J.}~\bibnamefont {Fan}}, \ and\ \bibinfo {author} {\bibfnamefont {G.}~\bibnamefont {Zeng}},\ }\href@noop {} {\bibfield  {journal} {\bibinfo  {journal} {Communications Physics}\ }\textbf {\bibinfo {volume} {7}},\ \bibinfo {pages} {276} (\bibinfo {year} {2024})}\BibitemShut {NoStop}%
\end{thebibliography}%

\end{document}